\documentclass[a4paper, 11pt, oneside]{article}

\usepackage{epsf,amsmath,amssymb,graphicx,dcolumn}
\usepackage{scalefnt,cite,hyperref,bbold}
\usepackage{cleveref,etoolbox,color,soul}
\usepackage{listings,booktabs}
\usepackage[title]{appendix}
\usepackage{chngcntr}
\usepackage{epstopdf}
\usepackage{mathrsfs}
\usepackage{dsfont}
\usepackage{amsfonts}

\counterwithin{equation}{section}

\newcommand{\half}{\tfrac{1}{2}}

\newlength{\absize}

\newenvironment{Eqnarray}%
     {\arraycolsep 0.14em\begin{eqnarray}}{\end{eqnarray}}
\def\beq{\begin{Eqnarray}}
\def\eeq{\end{Eqnarray}}
\def\beqa{\begin{Eqnarray}}
\def\eeqa{\end{Eqnarray}}
\def\bea{\begin{Eqnarray}}
\def\eea{\end{Eqnarray}}
\def\phm{\phantom{-}}

\newcommand{\dvec}[1]{\overset{\lower 1pt\hbox{\text{\scriptsize$\leftrightarrow$}}}{#1}}

\def\T{{\mathsf T}}
\renewcommand{\overrightarrow}[1]{\overset{\lower
    1pt\hbox{\text{\scriptsize$\rightarrow$}}}{#1}}
\renewcommand{\overleftarrow}[1]{\overset{\lower 1pt\hbox{\text{\scriptsize$\leftarrow$}}}{#1}}

\makeatletter
\let\save@mathaccent\mathaccent
\newcommand*\if@single[3]{%
  \setbox0\hbox{${\mathaccent"0362{#1}}^H$}%
  \setbox2\hbox{${\mathaccent"0362{\kern0pt#1}}^H$}%
  \ifdim\ht0=\ht2 #3\else #2\fi
  }
\newcommand*\rel@kern[1]{\kern#1\dimexpr\macc@kerna}
\newcommand*\widebar[1]{\@ifnextchar^{{\wide@bar{#1}{0}}}{\wide@bar{#1}{1}}}
\newcommand*\wide@bar[2]{\if@single{#1}{\wide@bar@{#1}{#2}{1}}{\wide@bar@{#1}{#2}{2}}}
\newcommand*\wide@bar@[3]{%
  \begingroup
  \def\mathaccent##1##2{%
    \let\mathaccent\save@mathaccent
    \if#32 \let\macc@nucleus\first@char \fi
    \setbox\z@\hbox{$\macc@style{\macc@nucleus}_{}$}%
    \setbox\tw@\hbox{$\macc@style{\macc@nucleus}{}_{}$}%
    \dimen@\wd\tw@
    \advance\dimen@-\wd\z@
    \divide\dimen@ 3
    \@tempdima\wd\tw@
    \advance\@tempdima-\scriptspace
    \divide\@tempdima 10
    \advance\dimen@-\@tempdima
    \ifdim\dimen@>\z@ \dimen@0pt\fi
    \rel@kern{0.6}\kern-\dimen@
    \if#31
      \overline{\rel@kern{-0.6}\kern\dimen@\macc@nucleus\rel@kern{0.4}\kern\dimen@}%
      \advance\dimen@0.4\dimexpr\macc@kerna
      \let\final@kern#2%
      \ifdim\dimen@<\z@ \let\final@kern1\fi
      \if\final@kern1 \kern-\dimen@\fi
    \else
      \overline{\rel@kern{-0.6}\kern\dimen@#1}%
    \fi
  }%
  \macc@depth\@ne
  \let\math@bgroup\@empty \let\math@egroup\macc@set@skewchar
  \mathsurround\z@ \frozen@everymath{\mathgroup\macc@group\relax}%
  \macc@set@skewchar\relax
  \let\mathaccentV\macc@nested@a
  \if#31
    \macc@nested@a\relax111{#1}%
  \else
    \def\gobble@till@marker##1\endmarker{}%
    \futurelet\first@char\gobble@till@marker#1\endmarker
    \ifcat\noexpand\first@char A\else
      \def\first@char{}%
    \fi
    \macc@nested@a\relax111{\first@char}%
  \fi
  \endgroup
}
\makeatother

\def\iso{\mathchoice{\cong}{\cong}{\isoS}{\cong}}
\def\isoS{\vbox{\baselineskip 0pt  \lineskip 0.5pt
    \ialign{$ \mathsurround=0pt  \scriptstyle \hfil ## \hfil $\crcr
        \sim \crcr = \crcr}}}

\let\Re\relax% Set equal to \relax so that LaTeX thinks it's not defined
\let\Im\relax% Set equal to \relax so that LaTeX thinks it's not defined
\DeclareMathOperator{\Re}{Re}
\DeclareMathOperator{\Im}{Im}

\def\nn{\nonumber}
\def\half{\tfrac{1}{2}}
\def\quarter{\tfrac{1}{4}}

\def\lam{\lambda}
\def\Lam{\Lambda}

\def\abar{{\bar a}}
\def\bbar{{\bar b}}
\def\cbar{{\bar c}}
\def\dbar{{\bar d}}
\def\ebar{{\bar e}}
\def\fbar{{\bar f}}
\def\gbar{{\bar g}}
\def\hb{{\bar h}}

\def\eq#1{eq.~(\ref{#1})}
\def\Eq#1{Equation~(\ref{#1})}

\def\eqs#1#2{eqs.~(\ref{#1}) and (\ref{#2})}

\def\eqss#1#2#3{eqs.~(\ref{#1}), (\ref{#2}), and (\ref{#3})}
\def\eqst#1#2{eqs.~(\ref{#1})--(\ref{#2})}

\def\lsub#1{_{\lower 1.5pt\hbox{$\scriptstyle#1$}}}
\def\lowsub#1{_{\lower 2.5pt\hbox{$\scriptstyle#1$}}}
\def\rsub#1{_{\raise 1.5pt\hbox{$\scriptstyle#1$}}}
\def\lsup#1{^{\lower 4pt\hbox{$\scriptstyle#1$}}}
\def\llsup#1{^{\lower 3pt\hbox{$\scriptstyle#1$}}}
\def\lowsup#1{^{\lower 2pt\hbox{$\scriptstyle#1$}}}
\def\rsup#1{^{\raise 1pt\hbox{$\scriptstyle#1$}}}
\def\rsuper#1{^{\raise 2pt\hbox{$\scriptstyle#1$}}}
\def\rrsup#1{^{\raise 4pt\hbox{$\scriptstyle#1$}}}
\def\lllsup#1{^{\lower 1pt\hbox{$\scriptstyle#1$}}}
\def\lwsup#1{^{\lower 0.25pt\hbox{$\scriptstyle#1$}}}

\def\newcdot{\kern.06em{\cdot}\kern.06em}

\newcounter{notecount}
\def\half{\tfrac12}

\begin{document}

\long\def\symbolfootnote[#1]#2{\begingroup%
\def\thefootnote{\fnsymbol{footnote}}\footnote[#1]{#2}\endgroup}

\vspace{0.1cm}

\begin{center}
\Large\bf\boldmath
RG-stable parameter relations of a scalar field theory in absence of a symmetry
\unboldmath
\end{center}
\vspace{0.05cm}
\begin{center}
Howard E.~Haber$^{a}$ and P. M.~Ferreira$^{b,c}$
\\[0.4cm]
{\small
{\sl${}^a$ Santa Cruz Institute for Particle Physics \\
 University of California, 1156 High Street, Santa Cruz, CA 95064 USA}\\[0.2em]
{\sl${}^b$Instituto  Superior  de  Engenharia  de  Lisboa,  Portugal}\\[0.2em]
{\sl${}^c$Centro  de  F{\'i}sica  Te{\'o}rica  e  Computacional,  Universidade  de  Lisboa,  Portugal}
}
\end{center}

\begin{abstract}

The stability of tree-level relations among the parameters of a quantum field theory with respect to
renormalization group (RG) running is typically explained by the existence of a symmetry.
We examine a toy model of a quantum field theory of two real scalars in which a tree-level relation
among the squared-mass parameters of the scalar potential appears to be RG-stable without
the presence of an appropriate underlying symmetry.   The stability of this relation with respect to
renormalization group running can be explained by complexifying the original
scalar field theory.  It is then possible to exhibit a symmetry that guarantees the relations of
relevant beta functions of squared-mass parameters of the complexified theory.
Among these relations, we can identify equations that are algebraically identical to the corresponding
equations that guarantee the stability of the relations among the squared-mass parameters of the original real scalar
 field theory where the symmetry of the complexified theory is no longer present.

\end{abstract}

\setcounter{footnote}{0}

\section{Introduction}
\label{intro}

The discovery of the Higgs boson at the LHC in 2012~\cite{ATLAS:2012yve,CMS:2012qbp}
provided strong evidence that the mechanism for generating the masses of the gauge bosons, quarks, and charged leptons of the Standard Model was governed by the dynamics of a weakly-coupled scalar sector.  Indeed, the Higgs boson appears to be
an elementary spin-0 particle, the first of
its kind. Subsequent measurements have shown that the Higgs boson couplings to fermions and gauge bosons
are, with increasing experimental precision~\cite{ATLAS:2022vkf,CMS:2022dwd}, nearly identical to those predicted
by the Standard Model (SM).
However, despite these impressive successes, a number of fundamental aspects of the theory of fundamental particles and their interactions remain unexplained.
As a result, the possibility of new physics beyond the SM has been
considered for decades.

Enlarging the
scalar sector beyond the one complex SU(2) doublet employed by the SM
has long been an interesting and promising
way to try to address some of the issues that the SM is incapable of explaining. For example,  a theory of very small but nonzero
neutrino masses may be achieved by considering an extra SU(2) triplet field,
using the see-saw mechanism~\cite{Mohapatra:1979ia}.
Adding gauge singlet scalars has been considered in order to generate a first order electroweak phase transition that is required for a viable theory of
electroweak baryogenesis~\cite{Barger:2008jx}.  Finally, adding a second complex scalar doublet to the SM, resulting in the two Higgs
doublet model (2HDM)~\cite{Lee:1973iz}, has been proposed as a means to explain dark matter~\cite{Deshpande:1977rw,Barbieri:2006dq},
or as a possible new source of CP-violation~\cite{Lee:1973iz}.  Even without a particular theoretical motivation, it is noteworthy that both the gauge sector and the fermion sector of the SM are quite nonminimal (as Rabi famously noted after the discovery of the muon by asking ``who ordered that?'').  Thus, it is certainly useful to entertain the possibility that the scalar sector of the SM should also be nonminimal.

Extending the scalar sector predicts the existence of new scalar particles and its attendant phenomenology. However, a larger scalar sector comes at a price.  Whereas
the SM scalar potential is fully characterized by two independent real parameters,
the scalar potential of an extended scalar sector introduces many additional parameters.
For example, the most general scalar potentials of the 2HDM and the three Higgs doublet model (3HDM) are governed by 14 and 54 real parameters, respectively.\footnote{To be more precise, the corresponding number of physical parameters is slightly less than the numbers quoted above after taking into account possible scalar field redefinitions~\cite{Davidson:2005cw}.  In particular, the 2HDM and 3HDM scalar sectors are governed by 11 and 46 real (physical) parameters, respectively~\cite{Nishi:2007nh,Bento:2017eti}.}
The increased number of parameters that govern the scalar potential significantly
 reduces the predictive power of extended Higgs sector models.

 One way to reduce the number of independent parameters of these models is to impose
 global symmetries, either discrete and/or continuous, as they eliminate or impose relations among the Lagrangian parameters.
Moreover, these symmetries are usually considered
 because they have interesting phenomenological consequences beyond simply reducing the dimensionality of the model parameter space.
 For example, by imposing a particular $\mathbb{Z}_2$ symmetry on the 2HDM Lagrangian~\cite{Hall:1981bc,Barger:1989fj,Aoki:2009ha}, one can ``naturally'' eliminate
 tree-level scalar-mediated flavor changing neutral currents (FCNCs) that otherwise would appear in the model.  In particular, the $\mathbb{Z}_2$ symmetry allows only one of the scalar doublets to couple to fermions of the same electric charge,
 and as a consequence the Yukawa interactions of the scalars to quarks and leptons are rendered flavor-diagonal~\cite{Glashow:1976nt,Paschos:1976ay}.   Moreover, the number of parameters
 of the $\mathbb{Z}_2$-symmetric 2HDM is reduced to seven due to the $\mathbb{Z}_2$ symmetry.  Note that one can still achieve flavor-diagonal Higgs-fermion couplings if the $\mathbb{Z}_2$ symmetry is softly broken, in which case the symmetry still imposes parameter relations among the dimension-four scalar self-coupling parameters at the expense of adding one additional squared-mass parameter to the model.

 Another example of a 2HDM symmetry is the  U(1) Peccei-Quinn symmetry~\cite{Peccei:1977ur}, which was initially introduced in an attempt to solve the strong QCD
problem. In total, there are six different global
symmetries~\cite{Ivanov:2005hg,Ivanov:2006yq,Ivanov:2007de,Ferreira:2009wh,Ferreira:2010yh,Haber:2021zva} one can impose on the
scalar sector of the SU(2)$_L\times$U(1)$_Y$ 2HDM.
These symmetries arise when imposing the invariance of the scalar potential under unitary field transformations that mix
both scalar doublets (so called Higgs-family symmetries) or their complex conjugates (so called generalized CP symmetries). In all cases cited above these
are unitary transformations that preserve the kinetic energy terms of the scalar doublets.\footnote{Additional symmetries of the scalar potential have also been considered in Refs.~\cite{Battye:2011jj,Pilaftsis:2011ed,BhupalDev:2014bir,Darvishi:2019dbh}
that are not preserved by the hypercharge U(1)$_Y$ interactions of the 2HDM.}

One well-known consequence of imposing a symmetry on a model is the fact that if a tree-level parameter relation, $X=0$, is the result of some symmetry $\cal{S}$, then
that parameter relation will be preserved to all orders of perturbation 
theory (e.g., see Ref.~\cite{Ma:2009ax}).
Note that if $\cal{S}$ is spontaneously broken, then there may be {\em finite} corrections
to $X=0$ that give it a non-zero value at some order of perturbation theory, but there will never be any {\em infinite} corrections to this relation.
Equivalently, if $X = 0$ due to a symmetry then the beta-function of $X$ obeys the same equation, $\beta_X = 0$, to all orders of perturbation theory.
That is, the parameter relation $X=0$ is stable with respect to renormalization group (RG) running.
One can extend this result in the case of a softly broken symmetry.  In particular, if there is a parameter relation, $X=0$, among the dimensionless parameters of the scalar potential, then $\beta_X=0$ to all orders in perturbation theory, since $\beta_X$ can only depend on the dimensionless parameters of the models, which respect the symmetry (whose breaking is due to parameters of the model with dimensions of mass to a positive power).

Suppose that the one-loop beta function $\beta_X=0$.  Does this imply the existence of a symmetry that imposes the tree-level condition $X=0$?   In general, the answer is no.  If one then computes the two-loop beta function $\beta_X$, one will generically find that it does not vanish if no symmetry exists to impose $X=0$.
 Recently, a curious result was discovered in the case of the 2HDM.  Denoting the two complex scalar doublets of the 2HDM by $\Phi_1$ and $\Phi_2$, the most general gauge-invariant renormalizable scalar potential is given by~\cite{Haber:1993an,Gunion:2002zf,Branco:2011iw}
\bea
V &=& m_{11}^2\Phi_1^\dagger\Phi_1+m_{22}^2\Phi_2^\dagger\Phi_2
-[m_{12}^2\Phi_1^\dagger\Phi_2+{\rm h.c.}]+\half\lambda_1(\Phi_1^\dagger\Phi_1)^2
+\half\lambda_2(\Phi_2^\dagger\Phi_2)^2
+\lambda_3(\Phi_1^\dagger\Phi_1)(\Phi_2^\dagger\Phi_2)\nonumber\\[8pt]
&&\quad
+\lambda_4(\Phi_1^\dagger\Phi_2)(\Phi_2^\dagger\Phi_1)
+\left\{\half\lambda_5(\Phi_1^\dagger\Phi_2)^2
+\big[\lambda_6(\Phi_1^\dagger\Phi_1)
+\lambda_7(\Phi_2^\dagger\Phi_2)\big]
\Phi_1^\dagger\Phi_2+{\rm h.c.}\right\}\,.
 \label{eq:pot}
\eea
In Ref.~\cite{Ferreira:2023dke} it was shown that the set of relations
\beq
m^2_{22} = - m^2_{11}\,,\qquad\quad \lambda_1 = \lambda_2\,,\qquad\quad \lambda_7 = -\lambda_6\,,
\label{eq:nsc}
\eeq
is a fixed point of the scalar sector parameter RG equations to all orders of perturbation theory.  That is, to all orders in the parameters of the scalar potential and neglecting the gauge and Yukawa couplings, one finds that
\beq \label{betaeqs}
\beta_{m^2_{22}} = - \beta_{m^2_{11}} \,,\qquad\quad \beta_{\lambda_1} = \beta_{\lambda_2}\,,\qquad\quad \beta_{\lambda_7} = -\beta_{\lambda_6}\,.
\eeq
Moreover, the beta function relations given in \eq{betaeqs} were shown to hold to all orders in the perturbation expansion when gauge interactions are included.  In particular, these relations still hold if Yukawa interactions are now taken into account up to two-loop order (which suggests but does not yet prove that the relations of \eq{betaeqs} remain valid to all orders in the perturbation expansion).
This result strongly suggested that some manner of symmetry is present in the model that
would explain the origin of the results obtained in \eq{betaeqs}.
However, whereas the relations among the quartic scalar self-couplings in \eq{eq:nsc} can be obtained
by imposing one of the six known global 2HDM symmetries~\cite{Ivanov:2005hg,Ivanov:2006yq,Ivanov:2007de,Ferreira:2009wh,Ferreira:2010yh,Haber:2021zva} (the symmetry usually denoted by GCP2), the relation $m^2_{22} = - m^2_{11}$ cannot be reproduced
by any of the known symmetries of the 2HDM.
Indeed, Ref.~\cite{Ferreira:2023dke} demonstrated that the parameter relation $m_{22}^2=-m_{11}^2$ cannot be the result of any symmetry consisting of  a scalar field transformation that is a unitary transformation
of both scalar doublets or their complex conjugates.

Ref.~\cite{Ferreira:2023dke} also showed that a formal way of obtaining the conditions of eq.~\eqref{eq:nsc}
is to write the 2HDM potential in terms of the four gauge invariant scalar-field bilinears $r_\mu$ ($\mu=0,1,2,3$) of Ref.~\cite{Ivanov:2006yq} (see also Refs.~\cite{Maniatis:2006fs,Maniatis:2007vn}) and require invariance
under the transformation $r_0\to -r_0$, where
$r_0 \equiv \half(\Phi_1^\dagger \Phi_1 + \Phi_2^\dagger \Phi_2)$. Clearly, there is no
unitary transformation of the two Higgs doublet fields, $\Phi_i\to\sum_i  U_{ij}\Phi_j$ (where $i,j\in\{1,2\}$), that yields $r_0 \rightarrow -r_0$.  Consequently an unconventional alternative was proposed.
After re-expressing the two complex doublet scalar fields in terms of real fields $\phi_i$ such that
\bea
\Phi_1=
\frac{1}{\sqrt{2}}\begin{pmatrix}
	\phi_1+i\phi_2 \\
	\phi_3+i\phi_4
\end{pmatrix}, \qquad\quad
\Phi_2=
\frac{1}{\sqrt{2}}\begin{pmatrix}
	\phi_5+i\phi_6 \\
	\phi_7+i\phi_8
\end{pmatrix},
\eea
the parameter relations exhibited in eq.~\eqref{eq:nsc} are a consequence of imposing invariance of the scalar potential under $r_0 \rightarrow -r_0$.\footnote{In Ref.~\cite{Pilaftsis:2024uub}, a covariant bilinear treatment of the one-loop 2HDM potential was developed, and it was shown
that the tree-level relations among parameters shown in \eq{eq:nsc} are broken by ultraviolet-finite corrections to the scalar potential.  
Indeed, if the transformation $r_0\to -r_0$ were a legitimate symmetry, then this symmetry must be spontaneously broken due to the nonzero 2HDM scalar field vacuum expectation values.
Consequently, the (spontaneously broken) symmetry permits only \textit{finite} radiative corrections to the tree-level parameter relations, as previously noted.}
Equivalently,
\begin{align}
\phi_1  &\rightarrow  i \phi_6\,, &  & & \phi_2  &\rightarrow  i \phi_5\,,  & & &
\phi_3 &\rightarrow  i \phi_8\,, &  & & \phi_4 &\rightarrow  i \phi_7\,,  \nonumber \\
\phi_5 &\rightarrow  -i \phi_2\,, & & & \phi_6 &\rightarrow  -i \phi_1\,,  &  & &
\phi_7 &\rightarrow  -i \phi_4\,, &  & & \phi_8 &\rightarrow  -i \phi_3\,. &
\label{eq:weird}
\end{align}

However, these transformations do not correspond to a legitimate symmetry transformation for two reasons.   First, the allowed symmetry transformations of real fields must involve real numbers, whereas the transformations of \eq{eq:weird} involve the imaginary number $i$.  This observation is also reflected by noting the transformations given by \eq{eq:weird} correspond to the following transformations of the complex doublet scalar fields,
\begin{alignat}{2}
\Phi_1&\rightarrow-\Phi_2^*\,, &\qquad\quad
\Phi_1^\dagger&\rightarrow \Phi_2^{\T},\nonumber\\
\Phi_2&\rightarrow \Phi_1^*, &\qquad\quad
\Phi_2^\dagger&\rightarrow -\Phi_1^{\T}.
\label{eq:trandou}
\end{alignat}
In particular, the transformation law of the complex conjugate field $\Phi_i^*$ is not the complex conjugate of the corresponding transformation law of $\Phi_i$.

The second problem with the proposed symmetry transformations of \eq{eq:weird} [or equivalently, \eq{eq:trandou}] is that these transformations reverse the sign of the kinetic energy terms of the scalar fields.  Ref.~\cite{Ferreira:2023dke} advanced the radical proposal where the spacetime coordinates themselves also transform via $x^\mu \rightarrow i x^\mu$. Equivalently, the covariant derivative must also transform as
$D_\mu\to iD_\mu$ (which implies that the gauge fields themselves must also similarly transform) in order that the kinetic energy terms of the scalar fields remain invariant.

The transformations proposed above, which collectively correspond to no known symmetry, were informally dubbed as ``GOOFy'' symmetries based on the names of the four authors
 of Ref.~\cite{Ferreira:2023dke}.  Whether they express something
deeper hitherto unknown in quantum field theory that can
provide a viable explanation of the all-orders fixed points of the beta functions to guarantee the RG-stability of the parameter relation $m_{22}^2=-m_{11}^2$ is
an open question.

In this paper, we shall propose a method for identifying a legitimate symmetry explanation for the origin of the parameter relation $m_{22}^2=-m_{11}^2$.  To simplify the argument, we shall examine a toy model of two real scalar fields that possesses an RG-stable parameter relation among the squared-mass parameters of the scalar potential which is of the same form as in the 2HDM example introduced above.  One could again try to invoke the GOOFy symmetries to explain the RG-stability of this parameter relation as in the 2HDM example above.  However, for the same reasons outlined above, we shall reject this proposal.

Instead, we will take inspiration from the process of complexification used in mathematics to create a complex vector space (or Lie algebra) starting from a real vector space (or Lie algebra).  Given a real scalar field theory, we can create a complex scalar field theory (called the complexified theory) by promoting the real scalar fields to complex scalar fields.  What looked like GOOFy symmetry transformations of the real scalar field theory are now legitimate symmetry transformations of the complexified theory.  Consequently, the parameter relations of the complexified theory are RG-stable.   For example, the complexification of the toy model of two real scalar fields will yield a complexified theory with the RG-stable parameter relation $m_{22}^2=-m_{11}^2$, corresponding to the relation of the corresponding beta functions,
$\beta_{m^2_{22}} = - \beta_{m^2_{11}}$ that is satisfied to all orders in perturbation theory.
 A careful analysis of these beta functions reveals a particular relation that is algebraically
 identical to the corresponding beta function relation of the original toy model of real scalar fields that guarantees the RG-stability of the parameter relation $m_{22}^2=-m_{11}^2$.   The end result is the RG-stability of the relation among the squared-mass parameters of the original real scalar field Lagrangian despite the fact that the symmetry of the complexified theory is no longer present in the original model.

 This paper is organized as follows.  In Section~\ref{sec:toy}, a toy model with two real scalar fields
 is presented that possesses an RG-invariant relation among the squared-mass parameters that is not guaranteed by
any legitimate symmetry.  In Section~\ref{sec:comp}, we introduce the notion of complexification
 of a scalar field theory, where each real scalar field is promoted to a complex scalar field and two symmetries of the complexified theory are imposed.   The first symmetry is chosen such that the holomorphic terms of the scalar potential of the complexified theory match precisely the corresponding terms that appear in the scalar potential of the original real scalar field theory.
 The second symmetry is a standard CP symmetry that imposes reality conditions on all scalar potential parameters of the complexified theory.
The one-loop beta functions of the complexified model are written out explicitly in
 Section~\ref{sec:beta}.  The vanishing of the appropriate combinations of one-loop and two-loop beta functions of the parameters of the complexified theory yield a set of equations.   In
 Section~\ref{sec:fixedpoints}, we show that a subset of these equations are algebraically identical to the corresponding beta function equations of the original theory of real scalar fields.  We argue that these arguments generalize to all orders in perturbation theory.  We thus conclude that the RG-stability of the parameter relations of the real scalar field Lagrangian is a consequence of symmetries of the complexified theory that are not present in the original real scalar field model.  In
 Section~\ref{sec:revis}, we outline a procedure for constructing additional examples of real scalar field theories with parameter relations whose RG-stability can only be explained by the existence of a symmetry of the complexified theory.
In  Section~\ref{sec:conc}, we recapitulate the main results obtained in this paper.  The implication of these results and their relation to the all-order
RG-stability of the 2HDM squared mass parameter relation described earlier in this section are outlined in Section~\ref{sec:future} along with some 
possible generalizations of this work.  Further details of our analysis have been relegated to three appendices.

\section{A toy model with RG-stable parameter relations in the absence of a symmetry}
\label{sec:toy}

Consider a quantum field theory of two real scalar fields $\varphi_1$ and $\varphi_2$, with the most general renormalizable Lagrangian
given by
\beq
\mathscr{L}=\partial_\mu\varphi_i\partial^\mu\varphi_i-\half m^2_{ij}\,\varphi_i\varphi_j\,-\frac{1}{4!}\,\lambda_{ijk\ell}\,\varphi_i\varphi_j \varphi_k\varphi_\ell\,,
\label{eq:VR}
\eeq
with real coefficients $m^2_{ij}$ and $\lambda_{ijk\ell}$ with $i,j,k,\ell\,\in\,\{1,2\}$, and an implied sum over repeated
indices.   In order to avoid terms linear and cubic in the fields, we have imposed a global parity symmetry $\varphi_1\to -\varphi_1$ and $\varphi_2\to- \varphi_2$
 (taken simultaneously).
Note that $m^2_{ij}=m^2_{ji}$ and thus there are three real degrees of freedom in the quadratic coefficients
($m_{11}^2$, $m_{22}^2$, and $m_{12}^2$).  Likewise, $\lambda_{ijk\ell}$ is completely symmetric with respect to permutations of its
indices, and thus yields five independent real degrees of freedom (conveniently chosen to be $\lambda_{1111}$, $\lambda_{1112}$, $\lambda_{1122}$, $\lambda_{1222}$, and $\lambda_{2222}$).

One can further reduce the number of free parameters of the theory by imposing an additional symmetry.   Note that the kinetic energy term in \eq{eq:VR} is invariant under
the symmetry transformation $\varphi_i\to Q_{ij}\varphi_j$ (with an implicit sum over $j$), where $Q$ is a $2\times 2$ real orthogonal matrix; i.e., $Q\in$~O(2).  Any conventional symmetry transformation that is being
considered to reduce the number of free parameters should be either O(2) or a (continuous or discrete) proper subgroup of O(2).

We now impose the following relations among the scalar potential parameters:
\beq \label{realconds}
m_{22}^2 = -m_{11}^2\,\qquad\quad\lambda_{1111}=\lambda_{2222}\,,\qquad\quad
\lambda_{1112}=-\lambda_{1222}\,.
\eeq
The corresponding scalar potential now takes the following form:
\beqa
&& \hspace{-0.1in}
V_R = \half m_{11}^2\,\left(\varphi_1^2 \,-\,\varphi_2^2\right)+
m_{12}^2\,\varphi_1\varphi_2
+\tfrac{1}{24}\lambda_{1111}\,\left(\varphi_1^4 +\varphi_2^4\right)
+\tfrac14\lambda_{1122}\,(\varphi_1\varphi_2)^2
+\tfrac16\lambda_{1112}\,\left(\varphi_1^2 \,-\,\varphi_2^2\right)\,\varphi_1\varphi_2\,,\nn\\
&& \phantom{line} \label{eq:pott}
\eeqa
where the subscript $R$ emphasizes that this is a theory of \textit{real} scalar fields.  In principle, one could choose to set $m_{12}^2=0$ by performing an appropriate change of scalar field basis, as discussed in Appendix~\ref{app:basis}. However, such a basis choice is not stable under RG running, so we choose to leave $m_{12}^2$ as a free parameter.

We now pose the following question: are the parameter relations exhibited in \eq{realconds} stable under RG running?   We can check this using the one-loop and two-loop beta functions given in the literature~\cite{Cheng:1973nv,Machacek:1984zw,Luo:2002ti,Schienbein:2018fsw,Sartore:2020pkk}.   Starting from the Lagrangian given by \eq{eq:VR}  and writing $\beta\equiv\beta^{I}+\beta^{II}$, the corresponding one-loop beta functions are given by
\beqa
\beta^{I}_{m_{ij}^2}&=&m^2_{mn}\lambda_{ijmn}\,, \label{betam2} \\
\beta^{I}_{\lambda_{ijk\ell}}&=&\frac18\sum_{\rm perm}\lambda_{ijmn}\lambda_{mnk\ell}=\lambda_{ijmn}\lambda_{mnk\ell}+\lambda_{ikmn}\lambda_{mnj\ell}+\lambda_{i\ell mn}\lambda_{mnjk}\,, \label{betalambda}
\eeqa
with an implicit sum over the repeated indices,
where $\sum_{\rm perm}$ in \eq{betalambda} denotes a sum over the permutations of the uncontracted indices, $i$, $j$, $k$, and $\ell$.
Likewise, the corresponding two-loop contributions to the beta functions are given by
\beqa
\beta^{II}_{m^2_{ij}}&=&\frac{1}{12}\bigl(\lambda_{ik\ell m}\lambda_{nk\ell m}m^2_{nj}+\lambda_{jk\ell m}\lambda_{nk\ell m}m^2_{ni}\bigr)-2m^2_{k\ell}\lambda_{ikmn}\lambda_{j\ell mn}\,, \label{twoloopbetam} \\
\beta^{II}_{\lambda_{ijk\ell}}&=&\frac{1}{72}\sum_{\rm perm}\lambda_{inpq}\lambda_{mnpq}\lambda_{mjk\ell}-\frac14\sum_{\rm perm}\lambda_{ijmn}\lambda_{kmpq}\lambda_{\ell npq}\,.
\label{twoloopbetalambda}
\eeqa

Using the results obtained in Appendix~\ref{app:beta}, we conclude that
\beqa
\beta_{m_{11}^2+m_{22}^2}\bigl|_{\rm sym}=\beta_{m_{11}^2}+\beta_{m_{22}^2}\bigl|_{\rm sym}&=&0\,,\label{betaone}\\
\beta_{\lambda_{1111}-\lambda_{2222}}\bigl|_{\rm sym}=\beta_{\lambda_{1111}}-\beta_{\lambda_{2222}}\bigl|_{\rm sym}&=&0\,,\label{betatwo}\\
\beta_{\lambda_{1112}+\lambda_{2221}}\bigl|_{\rm sym}=\beta_{\lambda_{1112}}-\beta_{\lambda_{2221}}\bigl|_{\rm sym}&=&0\,,\label{betathree}
\eeqa
at both one-loop and two-loop order,
where ``sym'' indicates that the parameter relations exhibited in \eq{realconds} have been applied in evaluating the corresponding beta functions given by the right-hand sides of
\eqst{betam2} {twoloopbetalambda}.  Note that the two-loop beta functions, $\beta^{II}_{m^2_{ij}}$ and $\beta^{II}_{\lambda_{ijk\ell}}$, each consist of the sum of two linearly independent combinations of tensor quantities.  Thus, each individual combination separately vanishes when the parameter relations exhibited in \eq{realconds} are applied,
as demonstrated in \eqss{aabb}{ccdd1}{ccdd2} of Appendix~\ref{app:beta}.
These results are not accidental, as it appears that \eqst{betaone}{betathree} are satisfied to all orders in perturbation theory.

One could understand the results obtained in \eqst{betaone}{betathree} if a symmetry could be identified that forces the scalar potential to take on the form exhibited in \eq{eq:pott}.
Consider the following symmetry transformation:
\beq
\varphi_1 \rightarrow \varphi_2\,,\qquad\quad \varphi_2 \rightarrow - \varphi_1\,.
\label{nogoof}
\eeq
Imposing this as a symmetry of the scalar potential yields
\beq \label{altrealconds}
m^2_{22} =m^2_{11}\,\qquad\quad m_{12}^2=0\,,\qquad\quad \lambda_{1111}=\lambda_{2222}\,,\qquad\quad
\lambda_{1112}=-\lambda_{1222}\,.
\eeq
Comparing with \eq{realconds}, we see that although the relations among the scalar self-couplings are the same, the relations among the squared-mass parameters are different.
However, the scalar self-coupling parameter relations must be RG-stable as these relations are a consequence of a softly-broken symmetry
(due to the fact that the beta functions for the $\lambda_{ijk\ell}$ are independent of the squared-mass parameters).  That is, \eqs{betatwo}{betathree}, to all orders
in perturbation theory, are a consequence of a softly broken symmetry.

Unfortunately, this argument does not explain why the squared mass relation, $m_{22}^2=-m_{11}^2$ is RG-stable.
Following Ref.~\cite{Ferreira:2023dke} and the discussion given in Section~\ref{intro} [see \eq{eq:weird}],
suppose we were to propose the following ``symmetry'' transformation,\footnote{This toy model and the corresponding ``symmetry'' were proposed in Ref.~\cite{BohdanLisbon} to study the validity of applying imaginary transformations
of real scalar fields and spacetime coordinates to the computation of the one-loop effective potential~\cite{Ferreira:2025ate}.}
\beq
\varphi_1 \rightarrow i \varphi_2\,,\qquad\quad \varphi_2 \rightarrow -i \varphi_1\,.
\label{eq:goof}
\eeq
If we were to require
that the general scalar potential is invariant with respect to \eq{eq:goof}, then $V_R$
would necessarily have the form shown in \eq{eq:pott}, where $m_{22}^2=-m_{11}^2$.
However, following the same arguments presented in Section~\ref{intro}, there are two serious problems with this proposal.  First, the symmetry corresponding to the transformation proposed in \eq{eq:goof} is not a subgroup of O(2).  Indeed, it simply does not make sense to
use non-real numbers in considering possible symmetry transformations of real scalar fields.
Second, even if one were to allow such a transformation,
the kinetic energy terms of the Lagrangian
change sign when the fields are transformed according to \eq{eq:goof}, whereas these terms should be invariant with respect to a legitimate symmetry transformation.
This is analogous to the result obtained by Ref~\cite{Ferreira:2023dke} when
applying the ``symmetry'' transformation [cf.~\eq{eq:weird}] of the 2HDM scalar potential given in \eq{eq:pot}.
As noted in Section~\ref{intro}, the authors of Ref.~\cite{Ferreira:2023dke} attempted to address this second problem above by extending the symmetry transformation to the spacetime
coordinates themselves, which affected the derivative that appears in the kinetic energy term such that the kinetic energy term was now invariant
with respect to the extended ``symmetry''.
But, as previously asserted,  this is not a legitimate symmetry transformation in any conventional sense.

Since \eq{eq:goof} is a transformation involving non-real numbers, perhaps it would be useful to rewrite the real scalar field theory with the scalar potential given by \eq{eq:pott} as the theory of a single complex field,
\beq \label{Phidef}
\Phi=\frac{\varphi_1+i\varphi_2}{\sqrt{2}}\,.
\eeq
Consider the Lagrangian,
\beq \label{complexLag}
\mathscr{L}=\partial_\mu\Phi\partial^\mu\Phi^*-m_1^2 \Phi^*\Phi-(m_2^2\Phi^2+{\rm c.c.})-\lambda_1(\Phi^*\Phi)^2-(\lambda_2\Phi^4+{\rm c.c.})-\bigl(\lambda_3\Phi^2+{\rm c.c.})\Phi^*\Phi\,,
\phantom{xxx}
\eeq
where ``c.c.'' stands for complex conjugate, and we have imposed the discrete symmetry $\Phi\to -\Phi$ to remove terms linear and cubic in the scalar fields.  \Eq{complexLag} is governed by three squared-mass terms ($m_1^2$, $\Re m_2^2$, $\Im m_2^2$) and five quartic couplings ($\lambda_1$, $\Re \lambda_2$, $\Im\lambda_2$,  $\Re \lambda_3$, $\Im\lambda_3$), where
$m_1^2$ and~$\lambda_1$ are real parameters.   Plugging in \eq{Phidef} into \eq{complexLag} and comparing with \eq{eq:VR}, it follows that
\beqa
m_1^2&=&\half(m_{11}^2+m_{22}^2)\,, \label{rel1}\\
m_2^2&=&\tfrac14(m_{11}^2-m_{22}^2-2i \,m_{12}^2)\,,\label{rel2}\\
\lambda_1&=&\tfrac{1}{16}\bigl(\lambda_{1111}+\lambda_{2222}+2\lambda_{1122}\bigr)\,,\label{rel3}\\
\lambda_2&=&\tfrac{1}{96}\bigl[\lambda_{1111}+\lambda_{2222}-6\lambda_{1122}-4i(\lambda_{1112}-\lambda_{1222})\bigr]\,,\label{rel4}\\
\lambda_3&=&\tfrac{1}{24}\bigl[\lambda_{1111}-\lambda_{2222}-2i(\lambda_{1112}+\lambda_{1222})\bigr]\,.\label{rel5}
\eeqa
If we now impose the parameter relations given in \eq{realconds}, it follows that $m_1^2=\lambda_3=0$.  As in \eq{eq:pott}, the resulting scalar potential of the complex scalar $\Phi$  is also governed by five real degrees of freedom ($\lambda_1$, $\Re\lambda_2$, $\Im\lambda_2$, $\Re m_2^2$, and $\Im m_2^2$):
\beq \label{eq:pott2}
V_R=(m_2^2\Phi^2+{\rm c.c.})+\lambda_1(\Phi^*\Phi)^2+(\lambda_2\Phi^4+{\rm c.c.})\,.
\eeq
Of course, the physical consequences of \eqs{eq:pott}{eq:pott2} are the same, as these are the same theories expressed in two different ways.
One can also check that
\beqa
\beta_{m_1^2}\bigl|_{\rm sym}&=&0\,, \label{betazero1} \\
\beta_{\lambda_3}\bigl|_{\rm sym}&=&0\,,\label{betazero2}
\eeq
where ``sym'' instructs one to set $m_1^2=\lambda_3=0$ when evaluating the corresponding beta functions.

In light of \eq{altrealconds}, consider the symmetry transformation,
\beq \label{altsym}
\Phi\to -i\Phi^*\,,
\eeq
which implies that $\Phi^*\to i\Phi$.  Applying this symmetry to \eq{complexLag} yields $m_2^2=\lambda_3=0$, whereas $m_1^2$ is a free parameter.
If we regard the symmetry exhibited by \eq{altsym} as a softly-broken symmetry of \eq{complexLag}, then this provides an explanation for \eq{betazero2} to all orders in perturbation theory,

Of course, the argument just given does not explain why the squared mass relation, $m_1^2=0$ is RG-stable.
Once again, we shall attempt to apply the ``symmetry'' transformation given by \eq{eq:goof}.  Rewriting this in terms of the complex field $\Phi$, we conclude that \eq{eq:goof} is equivalent to the ``symmetry'' transformation,
\beq \label{GOOFy}
\Phi\to\Phi\,,\qquad\quad \Phi^*\to -\Phi^*\,.
\eeq
Although this proposed symmetry transformation does indeed set $m_1^2=\lambda_3=0$, \eq{GOOFy} does not make sense as a symmetry transformation of a
complex scalar field theory since the transformation law for $\Phi^*$ is not the complex conjugate of the transformation law of $\Phi$.  Indeed, this result is analogous to the
``symmetry'' of the 2HDM scalar potential given in \eq{eq:pot} that was proposed in Ref.~\cite{Ferreira:2023dke} [cf.~eqs.~\eqref{eq:trandou}].
Moreover, the kinetic energy term
changes sign under \eq{GOOFy} as previously noted below \eq{eq:goof}.
Thus, \eq{GOOFy} cannot be used to explain the RG fixed point exhibited in \eq{betazero1}.  Of course, these two problems are the same ones noted when discussing the proposed symmetry transformation for the real scalar field theory above.

For these reasons, we shall reject the proposed extended GOOFy symmetry of Ref.~\cite{Ferreira:2023dke} as an explanation for the fixed-point
behaviors exhibited in \eqs{betaone}{betazero1}.
Indeed, any conventional symmetry that preserves the kinetic energy term will also preserve the term $m_1^2(\Phi^*\Phi)$ in \eq{complexLag}.  Hence, no
conventional symmetry can set $m_1^2=0$.

\section{Complexification of the toy model of two real scalar fields}
\label{sec:comp}

A symmetry transformation such as \eq{eq:goof} would make sense if the corresponding scalar fields were complex.   This motivates a procedure, which we denote by \textit{complexification},
where the scalar fields of the real scalar field theory are promoted to complex scalar fields denoted by $\Phi_i$.   When applied to the toy model of Section~\ref{sec:toy}, we can express the two complex scalar fields $\Phi_i$ ($i\in\{1,2\}$) in terms of four real scalar fields $\varphi_i$ where $i\in\{1,2,3,4\}$, as follows:
\beq \label{complexreal}
\Phi_1\,=\, \frac{1}{\sqrt{2}}\left(\varphi_1+i\,\varphi_2\right)\,,\qquad\quad
\Phi_{2}\,=\, \frac{1}{\sqrt{2}}\left(\varphi_3+i\,\varphi_4\right)\,.
\eeq
Moreover, the complexified model is \textit{defined} to employ a canonically normalized kinetic energy term,
\beq \label{KE}
\mathscr{L}_{\rm KE}= \partial^\mu\Phi^*_{\abar}\partial_\mu\Phi_a\,,
\eeq
which is invariant under the U(2) transformation,
\beq \label{youtwo}
\Phi_a\to U_{a\bbar}\Phi_b\,,\qquad \Phi^*_{\abar}\to \Phi^*_{\bbar}U^\dagger_{b\abar}\,,
\eeq
where $U^\dagger_{b\abar}U_{a\cbar}=\delta_{b\cbar}$.
In the above notation, the indices $a,  b, c\in\{1,2\}$ and
$\abar, \bbar, \cbar \in\{\bar{1},\bar{2}\}$ run over the complex two dimensional flavor space of scalar fields. The use of unbarred/barred index notation is accompanied by the rule that there is an implicit sum over unbarred/barred index pairs.

As in the original model of real scalar fields, we
shall impose a parity symmetry,
\beq
\Phi_1\to -\Phi_1 \quad \mbox{and} \quad\Phi_2\to -\Phi_2 \quad \mbox{(taken simultaneously)},
\label{eq:parit}
\eeq
 to remove terms in the scalar potential with an odd number of fields.   In this case,
the most general
renormalizable scalar potential of the complexified model may be written as
\bea
V_C &=&  M^2_{a\bbar} \Phi^*_{\abar} \Phi_b+M^2_{\abar\bbar} \Phi_a \Phi_b +M^2_{ab}\Phi^*_{\abar}\Phi^*_{\bbar}+\Lambda_{ab\cbar\dbar}\,\Phi^*_{\abar}\Phi^*_{\bbar}\Phi_c\Phi_d\nonumber\\[4pt]
&&
+ \Lambda_{\abar\bbar\cbar\dbar}\, \Phi_a \Phi_b \Phi_c \Phi_d
+\Lambda_{a\bbar \cbar \dbar} \,\Phi^*_{\abar} \Phi_b \Phi_c \Phi_d
+ \Lambda_{abc\dbar} \,\Phi^*_{\abar} \Phi^*_{\bbar}\Phi^*_{\cbar} \Phi_d
+\Lambda_{abcd} \Phi^*_{\abar} \Phi^*_{\bbar}\Phi^*_{\cbar} \Phi^*_{\dbar}\,,
\label{eq:VC}
\eea
where the subscript ``C" emphasizes that this is the complexified version of the original toy model of two real scalar fields.
In the notation used in \eq{eq:VC}, the squared-mass parameters $M^2_{ab}$ and $M^2_{a\bbar}$  are independent (despite the use of the same symbol $M^2$).  Likewise, the quartic coupling parameters $\Lambda_{abcd}$, $\Lambda_{abc\dbar}$, and $\Lambda_{ab\cbar\dbar}$ are independent (despite the use of the same symbol $\Lambda$).   One can distinguish among the independent parameters based on their explicit unbarred/barred index structure.

The squared-mass and  quartic coupling parameters satisfy the following relations:
\beq \label{indexrels}
M^2_{\abar\bbar}=M^2_{\bbar\abar}\,,\qquad M^2_{ab}=M^2_{ba}\,,\qquad \Lambda_{ab\cbar \dbar}=\Lambda_{ba\cbar\dbar}=\Lambda_{ab\dbar\cbar}=\Lambda_{ba\dbar\cbar}\,.
\eeq
Similarly, $\Lambda_{\abar\bbar\cbar\dbar}$ is symmetric under the permutation of the indices $\abar\bbar\cbar\dbar$,
$\Lambda_{a\bbar \cbar \dbar}$ is symmetric under the permutation of the indices $\bbar\cbar \dbar$,
$\Lambda_{ab\cbar\dbar}$ is separately symmetric under the interchange of the indices $ab$ and $\cbar\dbar$, respectively [as indicated in \eq{indexrels}],
$\Lambda_{abc\dbar}$ is symmetric under the permutation of the indices $abc$,
and $\Lambda_{abcd}$ is symmetric under the permutation of the indices $abcd$.

Hermiticity of $V_C$ implies that
\beq \label{reality}
M^2_{a\bbar}=\bigl[M^2_{b \abar}\bigr]^*\,,\qquad\Lambda_{ab \cbar\dbar} =\bigl[\Lambda_{cd\abar\bbar}\bigr]^*\,,
\eeq
and
\beq \label{spurions}
 M^2_{\abar\bbar} =\bigl[M_{ab}^2\bigr]^*\,,\qquad \Lambda_{\abar\bbar \cbar \dbar}=\bigl[\Lambda_{abcd}\bigr]^*\,,\qquad
\Lambda_{d\abar\bbar\cbar} =\bigl[\Lambda_{abc\dbar}\bigr]^*\,.
\eeq
In particular, $M^2_{1\bar{1}}$, $M^2_{2\bar{2}}$, $\Lambda_{11\bar{1}\bar{1}}$, $\Lambda_{22\bar{2}\bar{2}}$, and $\Lambda_{12\bar{1}\bar{2}}=\Lambda_{21\bar{2}\bar{1}}$ are real parameters.

We can now identify the independent parameters of $V_C$.   There are ten independent squared-mass parameters,
\beq \label{ten}
M^2_{1\bar{1}}, M^2_{2\bar{2}}, \Re M^2_{1\bar{2}}, \Im M^2_{1\bar{2}}, \Re M^2_{11}, \Re M^2_{12}, \Re M^2_{22},  \Im M^2_{11}, \Im M^2_{12}, \Im M^2_{22},
\eeq
and 35 independent quartic coupling parameters,
\beqa
&& \Re \Lambda_{1111},  \Re \Lambda_{1112},  \Re \Lambda_{1122},  \Re \Lambda_{1222}, \Re\Lambda_{2222},  \Im \Lambda_{1111},  \Im \Lambda_{1112},  \Im \Lambda_{1122},  \Im \Lambda_{1222}, \Im\Lambda_{2222},  \nn \\
&& \Re \Lambda_{111\bar{1}}, \Re \Lambda_{112\bar{1}}, \Re \Lambda_{122\bar{1}}, \Re \Lambda_{222\bar{1}},  \Re \Lambda_{111\bar{2}}, \Re \Lambda_{112\bar{2}}, \Re \Lambda_{122\bar{2}}, \Re \Lambda_{222\bar{2}},  \nn \\
&& \Im \Lambda_{111\bar{1}}, \Im \Lambda_{112\bar{1}}, \Im \Lambda_{122\bar{1}}, \Im \Lambda_{222\bar{1}},  \Im \Lambda_{111\bar{2}}, \Im \Lambda_{112\bar{2}}, \Im \Lambda_{122\bar{2}}, \Im \Lambda_{222\bar{2}},  \nn \\
&& \Lambda_{11\bar{1}\bar{1}}, \Lambda_{22\bar{2}\bar{2}}, \Lambda_{12\bar{1}\bar{2}}, \Re\Lambda_{11\bar{1}\bar{2}}, \Re\Lambda_{11\bar{2}\bar{2}}, \Re\Lambda_{12\bar{2}\bar{2}},
\Im\Lambda_{11\bar{1}\bar{2}}, \Im\Lambda_{11\bar{2}\bar{2}}, \Im\Lambda_{12\bar{2}\bar{2}}\,. \label{thirtyfive}
\eeqa

Although $V_C$ is \textit{not} invariant under a U(2) transformation exhibited in \eq{youtwo}, one can interpret \eq{youtwo} as a change in the scalar field basis.
The benefit of the unbarred/barred index notation is that the index structure of the scalar potential parameters indicates how these parameters change under a scalar field basis transformation:
\beqa
&&
M^2_{a\bbar}\to U_{a\cbar}U^\dagger_{d\bbar}M^2_{c\dbar}\,, \qquad   M^2_{ab}\to U_{a\cbar}U_{b\dbar}M^2_{cd}\,,\qquad \Lambda_{ab \cbar\dbar}\to U_{a\ebar}U_{b\fbar} U^\dagger_{g\cbar} U^\dagger_{h\dbar} \Lambda_{ef\gbar\hb}\,,\nn \\
&&
 \Lambda_{abcd}\to U_{a\ebar}U_{b\fbar}U_{c\gbar}U_{d\hb}\Lambda_{efgh}\,,\qquad  \Lambda_{abc\dbar}\to U_{a\ebar}U_{b\fbar}U_{c\gbar}U^\dagger_{h\dbar}\Lambda_{efg\hb}\,.
\eeqa
We shall now impose two additional symmetries to precisely define the complexification of the toy model of Section~\ref{sec:toy} [whose scalar potential is given by \eq{eq:pott}].
Possible symmetries to consider are any of the continuous or discrete subgroups of U(2), or generalized CP (GCP) transformations, $\Phi_a\to V_{ab}\Phi^*_\bbar$,
where $V$ is a fixed $2\times 2$ unitary matrix.  

First, we shall promote the illegitimate symmetry transformations exhibited in \eq{eq:goof} to a legitimate symmetry of the complexified model by 
requiring that the scalar potential shown in \eq{eq:VC} is invariant under\footnote{In terms of the $\varphi_i$ defined in \eq{complexreal}, the transformations of \eq{eq:sym} correspond to $\varphi_1\leftrightarrow -\varphi_4$ and $\varphi_2\leftrightarrow\varphi_3$.}
\beq
\Phi_1 \rightarrow \,i\,\Phi_2\;\;\; , \;\;\; \Phi_2 \rightarrow \,-i\,\Phi_1\,.
\label{eq:sym}
\eeq
Imposing \eq{eq:sym} as a symmetry of $V_C$ yields
\beqa
&&M_{1\bar{1}}^2=M_{2\bar{2}}^2\,,\qquad\qquad
\Re M^2_{1\bar{2}}=0\,,\qquad\qquad\,\,\, M_{11}^2=-M_{22}^2\,,\label{symcond1}\\
&&\Lambda_{1111}=\Lambda_{2222}\,,\qquad\quad\Lambda_{1112}=-\Lambda_{1222}\,,\label{symcond2}\\
&& \Lambda_{111\bar{1}}=-\Lambda_{222\bar{2}}\,,\qquad\,\, \Lambda_{112\bar{1}}=\Lambda_{122\bar{2}}\,,\qquad\quad\, \Lambda_{112\bar{2}}=-\Lambda_{122\bar{1}}\,,
\qquad\quad\Lambda_{111\bar{2}}=\Lambda_{222\bar{1}}\,,\label{symcond3}\\
&& \Lambda_{11\bar{1}\bar{1}}=\Lambda_{22\bar{2}\bar{2}}\,,\qquad\quad\, \Lambda_{11\bar{1}\bar{2}}=-\Lambda^*_{12\bar{2}\bar{2}}\,,\qquad\,\, \Lambda_{11\bar{2}\bar{2}}=\Lambda^*_{11\bar{2}\bar{2}}\,, \label{symcond4}
\eeqa
where in \eq{symcond4}, we have made use of the last relation of \eq{spurions}.  This leaves us with six independent squared-mass parameters and 19 independent quartic coupling parameters. Observe that the last relation in eq.~\eqref{symcond1} and the two relations of \eq{symcond2}
[which exclusively depend on self coupling tensors with four unbarred indices] are analogous to the three relations given in \eq{realconds}.

The scalar potential subject to the symmetry conditions given by \eqst{symcond1}{symcond4} is given by
\beqa
V_C&=& M^2\left( |\Phi_1|^2 \,+|\Phi_2|^2 \right)+i\Im M^2_{1\bar{2}}\bigl(\Phi_1^*\Phi_2-\Phi_1\Phi_2^*\bigr)
+\bigl[\bar{M}^2\left(\Phi_1^2- \Phi_2^2\right)+
 M^2_{12}\,\Phi_1 \Phi_2 + {\rm c.c.}\bigr]\nn \\
& & \!+  \Lambda_1\left( |\Phi_1|^4 + |\Phi_2|^4\right) +
\Lambda_2|\Phi_1|^2   |\Phi_2|^2  +
\left[\Lambda_3\left(\Phi_1^\ast \Phi_2\right)^2  + {\rm c.c.} \right]
+\left[\Lambda_4\,\Phi_1^\ast\,\Phi_2^\ast \left(\Phi_1^2 -  \Phi_2^2\right) + {\rm c.c.} \right]
\nonumber\\
&  &
+ \left[\Lambda_5 \left(\Phi_1  \Phi_2\right)^2  + {\rm c.c.} \right]+
\left[\Lambda_6\bigl( \Phi_1^4 +\Phi_2^4\bigr)+ {\rm c.c.} \right]+\left[\Lambda_7\Phi_1\,\Phi_2 \left(\Phi_1^2 -  \Phi_2^2\right) + {\rm c.c.} \right]
\nonumber\\
&  &
+ \bigl(\Lambda_8\,\Phi_1\,\Phi_2 + {\rm c.c.} \bigr)\left(|\Phi_1|^2 + |\Phi_2|^2\right)
+\bigl[\Lambda_9(\Phi_1^2|\Phi_1|^2-\Phi^2_2|\Phi_2|^2)+{\rm c.c.}\bigr]  \nonumber\\[5pt]
& &
+ \bigl[\Lambda_{10}(\Phi_1^2|\Phi_2|^2-\Phi^2_2|\Phi_1|^2)+{\rm c.c.}\bigr]
+\bigl[\Lambda_{11}\bigl(\Phi_1^3\Phi_2^*+\Phi_2^3\Phi_1^*\bigr)+{\rm c.c.}\bigr]\,,\label{V2plusV4}
\eeqa
where $M^2\equiv M^2_{1\bar{1}}=M^2_{2\bar{2}}$ and $\bar{M}^2\equiv M_{11}^2=-M^2_{22}$.

Second, we shall require that $V_C$ is invariant under a ``standard" CP transformation,
\beq
\Phi_1 \rightarrow \Phi_1^\ast \;\;\; , \;\;\; \Phi_2 \rightarrow \Phi_2^\ast\,,
\label{eq:CP}
\eeq
so that all scalar potential coefficients are real.
It then follows that $M^2_{1\bar{2}}=0$,
which finally leaves us with three independent real squared-mass
parameters ($M^2$, $\bar{M}^2$, and $M_{12}^2$) and 11 independent real quartic coupling parameters ($\Lambda_i$ for
$i=1,2,\ldots,11$) that govern the complexification of the toy model of Section~\ref{sec:toy}.  In particular, note that
\beqa
M^2_{ab}&=&M^2_{\abar\bbar}\ni \{\bar{M}^2,M^2_{12}\}\,,\\
M^2_{a\bbar}&=&M^2_{b\abar}\ni \{M^2\}\,,\\
\Lambda_{ab\cbar\dbar}&=&\Lambda_{cd\abar\bbar}\ni \{\Lambda_1,\Lambda_2,\Lambda_3,\Lambda_4\}\,,\\
\Lambda_{abcd}&=&\Lambda_{\abar\bbar\cbar\dbar}\ni \{\Lambda_5,\Lambda_6,\Lambda_7\}\,,\\
\Lambda_{abc\dbar}&=&\Lambda_{a\bbar\cbar\dbar}\ni \{\Lambda_8,\Lambda_9,\Lambda_{10},\Lambda_{11}\}\,,\label{linind}
\eeqa
after making use of \eqs{reality}{spurions}.
Since $M^2$, $\bar{M}^2$, and $M_{12}^2$ are independent real parameters, it follows that they are linearly independent, which implies that $M^2_{ab}$ and $M^2_{a\bbar}$ are also linearly independent.
Similarly, the $\Lambda_i$ are linearly independent real parameters, which implies that
$\Lambda_{ab\cbar\dbar}$, $\Lambda_{abcd}$, and $\Lambda_{abc\dbar}$ are also linearly independent. This is a crucial observation for the method we will propose
to explain the RG stability of parameter relations observed in the original toy model of two real scalar fields.

The complexification of the toy model of two real scalar fields with scalar potential given by
\eq{eq:pott} has been achieved by promoting the GOOFy symmetry of the toy model to a legitimate symmetry of the complexified model.   In particular, it is instructive to examine the terms of $V_C$ given in \eq{V2plusV4} that are holomorphic in the complex fields (i.e., those terms that depend just on the fields $\Phi_1$ and $\Phi_2$ but not their complex conjugates):
\beq \label{Vcholo}
V_C\ni \bar{M}^2\left(\Phi_1^2- \Phi_2^2\right)+
 M^2_{12}\,\Phi_1 \Phi_2 +\Lambda_5 \left(\Phi_1  \Phi_2\right)^2 +
\Lambda_6\bigl( \Phi_1^4 +\Phi_2^4\bigr)+\Lambda_7\Phi_1\,\Phi_2 \left(\Phi_1^2 -  \Phi_2^2\right)\,,
\eeq
where $\bar{M}^2$, $M_{12}^2$, $\lambda_5$, $\lambda_6$, and $\Lambda_7$ are real parameters.  It is noteworthy that \eq{Vcholo} has precisely the same form as \eq{eq:pott}.
This indicates that the complexification of the toy model of two real scalar fields has been properly obtained.

\section{One-loop beta functions of the complexified model}
\label{sec:beta}

First, let us  re-express the complex fields $\Phi_1$ and $\Phi_2$ in terms of the four real fields $\varphi_i$ ($i=1,2,3,4$) using \eq{complexreal}.  We can then rewrite the Lagrangian of two complex scalar fields given by \eqs{KE}{eq:VC} in the form shown in \eq{eq:VR}.  We shall call this process \textit{realification}.
Of course, the theory of two complex fields and the corresponding realified theory of four real scalar fields are the same model written in a different form.%
\footnote{We have already noted in Section~\ref{sec:toy} that the realification of a theory of a single complex field yields the most general theory of two real scalar fields.}
We can now make use of the results of Refs.~\cite{Cheng:1973nv,Machacek:1984zw,Luo:2002ti,Schienbein:2018fsw,Sartore:2020pkk} to
evaluate the beta functions of the squared-mass and quartic coupling parameters. In particular, starting from a theory written in terms of real scalar fields with a Lagrangian given by \eq{eq:VR}, the corresponding one-loop beta functions are given by \eqs{betam2}{betalambda}.

We begin with the
squared mass parameters.  Using the results of Appendix~\ref{app1}, one can
solve for the $M^2_{ab}$ and $M^2_{a\bbar}$ in terms of the $m_{ij}^2$, where $i,j\in\{1,2,3,4\}$.  For example,
\beqa
\Re M^2_{11}&=&\quarter\bigl(m_{11}^2-m_{22}^2\bigr)\,,\\
\Re M^2_{22}&=&\quarter\bigl(m_{33}^2-m_{44}^2\bigr)\,,\\
\Im M_{11}^2&=&\half m_{12}^2\,,\\
\Im M_{22}^2&=& \half m_{34}^2\,,
\eeqa
prior to imposing the symmetries specified in \eqs{eq:sym}{eq:CP}.
These results can be used to obtain the beta functions of the parameters $M^2_{\abar\bbar}$ and $M^2_{a\bbar}$.
For example, in light of \eq{symcond1} and the reality of all scalar potential parameters, it follows that
\beq
\beta_{M_{11}^2+M_{22}^2}\Bigl|_{\rm sym}=\quarter\left[\beta_{m_{11}^2}-\beta_{m_{22}^2}+\beta_{m_{33}^2}-\beta_{m_{44}^2}\right]\Bigl|_{\rm sym}=0\,,
\eeq
after imposing the relevant parameter relations (as indicated by the subscript ``sym'').
Of course, these results must hold to all orders
in perturbation theory as they are guaranteed by the symmetries of the complexified theory given by \eq{V2plusV4}.

For our purposes, it is more useful to re-express the one-loop beta functions of the quadratic parameters, given in \eq{betam2}, directly in terms of the parameters exhibited in \eqs{ten}{thirtyfive} that govern the complexified theory.   A straightforward calculation yields
\beqa
\beta^{I}_{M^2_{\abar\bbar}}&=& 4M_{\cbar\dbar}^2\Lambda_{cd\abar\bbar}+24M_{cd}^2\Lambda_{\abar\bbar\cbar\dbar}+6M^2_{e\dbar}\Lambda_{d\abar\bbar\ebar}\,,\label{betaMsq1}\\
\beta^{I}_{M^2_{a\bbar}}&=&12M^2_{cd}\Lambda_{a\bbar\cbar\dbar}+12 M^2_{\cbar\dbar}\Lambda_{acd\bbar}+8M^2_{d\ebar}\Lambda_{ae\bbar\dbar}\,.\label{betaMsq2}
\eeqa

We next consider the quartic coupling parameters.
Using the results of Appendix~\ref{app1}, one can now solve for $\Lambda_{abcd}$, $\Lambda_{abc\dbar}$ and $\Lambda_{ab\cbar\dbar}$ in terms of the $\lambda_{ijk\ell}$.  For example,
\begingroup
\allowdisplaybreaks
\beq
\Re\Lambda_{1111}&=&\frac{1}{96}\bigl[\lam_{1111}+\lambda_{2222}-6\lam_{1122}\bigr]\,,\\
\Re\Lambda_{2222}&=&\frac{1}{96}\bigl[\lam_{3333}+\lambda_{4444}-6\lam_{3344}\bigr]\,,\\
\Re\Lambda_{1112}&=&\frac{1}{96}\bigl[\lam_{1113}+\lambda_{2224}-3(\lam_{1124}+\lam_{1223})\bigr]\,,\\
\Re\Lambda_{1222}&=&\frac{1}{96}\bigl[\lam_{1333}+\lambda_{2444}-3(\lam_{1344}+\lam_{2334})\bigr]\,,\\
\Im\Lambda_{1111}&=&\frac{1}{24}\bigl(\lambda_{1112}-\lambda_{1222}\bigr)\,,\\
\Im\Lambda_{2222}&=&\frac{1}{24}\bigl(\lambda_{3334}-\lambda_{3444}\bigr)\,,\\
\Im\Lambda_{1112}&=&\frac{1}{96}\bigl[\lambda_{1114}-\lambda_{2223}+3(\lambda_{1123}-\lambda_{1224})\bigr]\,,\\
\Im\Lambda_{1222}&=&\frac{1}{96}\bigl[\lambda_{2333}-\lambda_{1444}+3(\lambda_{1334}-\lambda_{2344})\bigr]\,,
\eeq
\endgroup
prior to imposing the symmetries specified in \eqs{eq:sym}{eq:CP}.

In light of \eq{symcond2} and the reality of all scalar potential parameters, it follows that
\beqa
\beta_{\Lambda_{1111}-\Lambda_{2222}}\bigl|_{\rm sym}&=&\beta_{\lam_{1111}+\lambda_{2222}-6\lam_{1122}-\lam_{3333}-\lambda_{4444}+6\lam_{3344}}\bigl|_{\rm sym}= 0\,,\\
\beta_{\Lambda_{1112}+\Lambda_{1222}}\bigl|_{\rm sym}&=&\beta_{\lam_{1113}+\lambda_{2224}-3\lam_{1124}-3\lam_{1223}+\lam_{1333}+\lambda_{2444}-3\lam_{1344}-3\lam_{2334}}\bigl|_{\rm sym}=0\,,
\eeqa
after imposing the relevant parameter relations.
Of course, these results hold to all orders
in perturbation theory as they are guaranteed by the symmetries imposed on $V_C$.

It is again more useful to re-express the one-loop beta functions of the quartic couplings, given in \eq{betalambda}, in terms of the independent parameters given
in \eq{thirtyfive}  that govern the complexified theory.
Another straightforward calculation yields:
\beqa
\beta^{I}_{\Lambda_{\abar\bbar\cbar\dbar}}&=&\sum_{\rm perm}\Lambda_{\abar\bbar\ebar\fbar}\Lambda_{ef\cbar\dbar}+\frac{3}{8}\sum_{\rm perm}\Lambda_{e\abar\bbar\fbar}\Lambda_{f\ebar\cbar\dbar} \nn \\
&=& 8(\Lambda_{\abar\bbar\ebar\fbar}\Lambda_{ef\cbar\dbar}+\Lambda_{\abar\cbar\ebar\fbar}\Lambda_{ef\bbar\dbar}+\Lambda_{\abar\dbar\ebar\fbar}\Lambda_{ef\bbar\cbar})+
3(\Lambda_{e\abar\bbar\fbar}\Lambda_{f\ebar\cbar\dbar}+\Lambda_{e\abar\cbar\fbar}\Lambda_{f\ebar\bbar\dbar}+\Lambda_{e\abar\dbar\fbar}\Lambda_{f\ebar\bbar\cbar}),\phantom{xxxx}
\label{betaLamabcd} \\
\beta^{I}_{\Lambda_{ab\cbar\dbar}}&=&144\Lambda_{abef}\Lambda_{\ebar\fbar\cbar\dbar}+4\bigl(\Lambda_{ab\ebar\fbar}\Lambda_{ef\cbar\dbar}+\Lambda_{ae\fbar\cbar}\Lambda_{bf\ebar\dbar}+\Lambda_{be\fbar\cbar}\Lambda_{af\ebar\dbar}+\Lambda_{ae\fbar\dbar}\Lambda_{bf\ebar\cbar}+\Lambda_{be\fbar\dbar}\Lambda_{af\ebar\cbar}\bigr) \nn \\
&& \qquad +9\bigl(2\Lambda_{abe\fbar}\Lambda_{f\ebar\cbar\dbar}+\Lambda_{aef\cbar}\Lambda_{b\ebar\fbar\dbar}+\Lambda_{bef\cbar}\Lambda_{a\ebar\fbar\dbar}+\Lambda_{aef\dbar}\Lambda_{b\ebar\fbar\cbar}+\Lambda_{bef\dbar}\Lambda_{a\ebar\fbar\cbar}\bigr)\,,\label{betaLamabcd2}\\
\beta^{I}_{\Lambda_{abc\dbar}}&=&4\bigl(\Lambda_{ab\ebar\fbar}\Lambda_{efc\dbar}+\Lambda_{ac\ebar\fbar}\Lambda_{efb\dbar}+
\Lambda_{bc\ebar\fbar}\Lambda_{efa\dbar}\bigr)+
8\bigl(\Lambda_{abe\fbar}\Lambda_{cf\ebar\dbar}+\Lambda_{ace\fbar}\Lambda_{bf\ebar\dbar}+\Lambda_{bce\fbar}\Lambda_{af\ebar\dbar}\bigr) \nn \\
&& \qquad + 24\bigl(\Lambda_{abef}\Lambda_{c\ebar\fbar\dbar}+\Lambda_{acef}\Lambda_{b\ebar\fbar\dbar}+\Lambda_{bcef}\Lambda_{a\ebar\fbar\dbar}\bigr)\,.\label{betaLamabcd3}
\eeqa
As a simple check of the expressions for the one-loop beta functions obtained above, suppose that we require that $V_C$ is invariant with respect to the U(1) symmetry where $\Phi_a\to e^{i\theta}\Phi_a$
(for $a=1,2)$. This would imply that $M^2_{ab}=\Lambda_{abcf}=\Lambda_{abc\dbar}=0$ in \eq{eq:VC}.  The beta functions exhibited in
\eqss{betaMsq1}{betaLamabcd} {betaLamabcd3} then yield
$\beta_{M^2_{\abar\bbar}}=\beta_{\Lambda_{\abar\bbar\cbar\dbar}}=\beta_{\Lambda_{abc\dbar}}=0$ at one loop order (after imposing the relevant parameter relations) as expected, and these relations must hold
to all orders of perturbation theory.

The structure of the expressions for the beta functions obtained in eqs.~(\ref{betaMsq1}), (\ref{betaMsq2}), and (\ref{betaLamabcd})--(\ref{betaLamabcd3}) is noteworthy.  In particular,
covariance with respect to the unbarred/barred indices is respected by the beta function equations.  That is, after carrying out the implicit summation over unbarred/barred index pairs, the index structure
the beta function must match the corresponding free indices appearing on the right-hand side of the beta function expression.  For example, the free indices on both sides of \eq{betaLamabcd} consist of four barred indices.
The covariance of the other beta function equations can be easily verified.

\section{RG-stability of scalar potential parameter relations guaranteed by symmetries of the complexified model}
\label{sec:fixedpoints}

The goal of this section is to show that RG-stability of the parameter relations of the complexified theory given in \eqst{symcond1}{symcond4}, which are guaranteed by the invariance of \eq{eq:VC}
under the symmetry transformations given by \eqs{eq:sym}{eq:CP}, implies the RG-stability of parameter relations [\eq{realconds}] of the scalar potential of the original toy model of two real scalar fields [\eq{eq:pott}].  As discussed in Section~\ref{sec:toy}, only one of the three parameter relations of the toy model ($m_{22}^2=-m_{11}^2$) cannot be explained by a legitimate symmetry within the framework of the original model of two real scalar fields.  Thus, we only need to examine the beta functions of the squared-mass parameters to achieve our goal.

Consider a linear relation on the parameters $m_{ij}^2$ of the toy model of two real scalar fields of the form $c_{ij}m^2_{ij}=0$ (with an implicit sum over repeated indices).  In Section~\ref{sec:toy}, we found that the corresponding one-loop beta function,
\beq \label{originalbetacond}
\beta^{I}_{c_{ij}m^2_{ij}}\bigl|_{\rm sym}=c_{ij}m^2_{k\ell}\lambda_{ijk\ell}\bigl|_{\rm sym}=0\,,
\eeq
with $c_{11}=c_{22}=1$ and $c_{12}=c_{21}=0$, where ``sym'' indicates that the parameter relations given by \eq{altrealconds} have been applied.  As a result, the relation $m_{22}^2=-m_{11}^2$ is RG-stable at one-loop order despite the absence of a symmetry to enforce the relation among squared-mass parameters.

For the complexified theory, we have identified a legitimate symmetry that
imposes a linear relation on the parameters $M^2_{\abar\bbar}$ of the form $c_{ab}M^2_{\abar\bbar}=0$, with $c_{11}=c_{22}=1$ and $c_{12}=c_{21}=0$ as before.   Then, the corresponding beta function, $\beta_{c_{ab}M^2_{\abar\bbar}}$, must vanish to all orders in perturbation theory.   Applying this relation to \eq{betaMsq1}, we note that the symmetry will also impose \textit{separate} independent relations among the product of scalar potential parameters $M_{\cbar\dbar}^2\Lambda_{cd\abar\bbar}$, $M_{cd}^2\Lambda_{\abar\bbar\cbar\dbar}$, and $M^2_{e\dbar}\Lambda_{d\abar\bbar\ebar}$, since these are linearly independent quantities as noted below \eq{linind}.  Hence, one may conclude that three separate relations must be satisfied:
\beqa
c_{ab}M_{\cbar\dbar}^2\Lambda_{cd\abar\bbar}\bigl|_{\rm sym}&=& 0\,, \label{sym1}\\
c_{ab}M_{cd}^2\Lambda_{\abar\bbar\cbar\dbar}\bigl|_{s\rm ym}&=& 0\,,\label{sym2}\\
c_{ab}M^2_{e\dbar}\Lambda_{d\abar\bbar\ebar}\bigl|_{\rm sym}&=& 0\,,\label{sym3}
\eeqa
where there is implied summation over unbarred/barred index pairs, and ``sym''
indicates that the symmetry relations satisfied by $M^2_{\cbar\dbar}$, $M_{\cbar d}$, $\Lambda_{cd\abar\bbar}$, $\Lambda_{\abar\bbar\cbar\dbar}$,and $\Lambda_{d\abar\bbar\ebar}$
[exhibited in \eqst{symcond1}{symcond4}] have been imposed.
In light of  the CP symmetry, which enforces all scalar potential parameters of $V_C$ to be real, we recognize \eqs{originalbetacond}{sym2} as being algebraically equivalent.
Thus, we  have understood \eq{originalbetacond} as a consequence of a symmetry of the complexified model.

This is not an accident of the one-loop beta functions.   Consider the corresponding two-loop beta function for the $m^2_{ij}$ given in \eq{twoloopbetam}.
Applying this result to the toy model of two real scalar fields subject to the conditions specified in \eq{realconds}, the following two conditions separately hold:
\beqa
c_{ij}\bigl(\lambda_{ik\ell m}\lambda_{nk\ell m}m^2_{nj}+\lambda_{jk\ell m}\lambda_{nk\ell m}m^2_{ni}\bigr)\bigl|_{\rm sym}&=&0\,,\label{twoloopr1}\\
c_{ij}m^2_{k\ell}\lambda_{ikmn}\lambda_{j\ell mn}\bigl|_{\rm sym}&=&0\,.\label{twoloopr2}
\eeqa
Consequently $\beta^{II}_{c_{ij}m^2_{ij}}=0$ despite the absence of a real symmetry imposed on the scalar Lagrangian.

Following our one-loop analysis, we shall consider the two-loop beta function for the complexified model.  If we follow our previous technique, we should rewrite \eq{twoloopbetam} in terms of the parameters $M^2_{\cbar\dbar}$, $M_{\cbar d}$, $\Lambda_{cd\abar\bbar}$, $\Lambda_{\abar\bbar\cbar\dbar}$, and $\Lambda_{d\abar\bbar\ebar}$.  However, we can identify the possible index structures of the various terms that respect the covariance properties of the corresponding beta functions.   Similar to \eqst{sym1}{sym3}, one can derive separate relations that must be satisfied if the beta function vanishes.  Since the complexified theory does possess a symmetry that guarantees the relations exhibited in \eqst{symcond1}{symcond4},
we are assured that the two-loop beta function will vanish.  Among all the relations obtained, we find that
\beqa
c_{ab}\bigl(\Lambda_{\abar\dbar\ebar\fbar}\Lambda_{cdef}M^2_{\cbar\bbar}+\Lambda_{\bbar\dbar\ebar\fbar}\Lambda_{cdef}M^2_{\cbar\abar}\bigr)\bigl|_{\rm sym}&=&0\,,\label{twoloopm1}\\
c_{ab}M^2_{cd}\Lambda_{\abar\cbar\ebar\fbar}\Lambda_{ef\dbar\,\bbar}\bigl|_{\rm sym}&=&0\,.\label{twoloopm2}
\eeqa
Since we have also imposed CP conservation, it follows that all the quartic couplings in \eqs{twoloopm1}{twoloopm2} are real.
Moreover, since $\Lambda_{\abar\bbar\cbar\dbar}$ and  $\Lambda_{ab\cbar\dbar}$ are independent,
then \eq{twoloopm2} must hold if we numerically set $\Lambda_{ab\cbar\dbar}=\Lambda_{\abar\bbar\cbar\dbar}$ in \eq{twoloopm2}.   This numerical procedure is consistent with the symmetry conditions specified in \eqs{symcond2}{symcond4}.\footnote{First, we set $\Lambda_{11\bar{2}\bar{2}}=\Lambda_{12\bar{1}\bar{2}}$ [i.e., $\Lambda_2=\Lambda_3$ in the notation of \eq{V2plusV4}]. It then follows that $\Lambda_{ab\cbar\dbar}$ is now a completely symmetric real tensor that satisfies the same symmetry conditions as $\Lambda_{\abar\bbar\cbar\dbar}$.  Thus, if \eq{sym4} is valid for arbitrary $\Lambda_{\abar\bbar\cbar\dbar}$ and $\Lambda_{ab\cbar\dbar}$, then this equation must continue to be valid if $\Lambda_{ab\cbar\dbar}$ is replaced by $\Lambda_{\abar\bbar\cbar\dbar}$.}
We can therefore conclude that \eqs{twoloopr1}{twoloopm1} are algebraically identical.
Likewise, \eqs{twoloopr2}{twoloopm2} are algebraically identical.  Thus, we have again explained the vanishing of the beta functions in the real scalar model as a consequence of a symmetry of the corresponding complexified model.  This conclusion can be extended to arbitrary loops.  In particular, there will always be an equation obtained in the complexified model 
that only involves tensors with an even number of unbarred and barred indices, respectively,  which is algebraically identical to a corresponding equation obtained in  the toy model of two real scalar fields.

For example, one can repeat the analysis for the three-loop beta functions using the results in the literature~\cite{Steudtner:2020tzo}, but the end result is the same.  One can always find expressions that are products of
$M^2_{\abar\bbar}$, $\Lambda_{\abar\bbar\cbar\dbar}$, $\Lambda_{ab\cbar\dbar}$ and their complex conjugates (the latter need not be distinguished as all squared-mass and quartic coupling parameters are real due to the CP symmetry).  Once the relevant relations have been found for the parameters of the complexified model, one can numerically set $\Lambda_{ab\cbar\dbar}=\Lambda_{\abar\bbar\cbar\dbar}$, if needed, as shown above to produce relations that are algebraically equivalent to the corresponding relations of the original real scalar field model. We stress that this relation between the quartic couplings is {\em not} a requirement of further symmetry of the model, but is merely a numerical
choice. Since the $\Lambda_{ab\cbar\dbar}$ and $\Lambda_{\abar\bbar\cbar\dbar}$ are independent tensors [subject to the parameter relations specified in \eqs{symcond2}{symcond4}], the relations we obtained above are valid for {\em any} values they
might take, which of course includes the case where these two tensors are taken to be equal.

As noted at the beginning of this section, we do not need to justify the RG-stability of the parameter relations among the scalar self-couplings of the original toy model of real scalar fields, since we successfully identified a softly-broken symmetry to account for the observed behavior of the corresponding beta functions.  Nevertheless, it is instructive to show that the symmetry of the complexified model specified in \eqst{symcond1}{symcond4} can also be used to establish the RG-stability of the parameter relations among the scalar self-couplings of the original toy model of real scalar fields.\footnote{It is possible that examples of real scalar field theory models exist that possess scalar coupling relations whose RG-stability cannot be explained by a symmetry.  In such cases, one would show using the methods of this section that the corresponding RG-stability of the scalar coupling parameter relations of the complexified theory imply the RG-stability of the coupling parameter relations of the original real scalar field theory.}

Suppose that a symmetry imposes a linear relation on the parameters $\Lambda_{\abar\bbar\cbar\dbar}$, $\Lambda_{ab\cbar\dbar}$, $\Lambda_{abc\dbar}$ of the form
\beq \label{linearconds}
 c_{abcd}\Lambda_{\abar\bbar\cbar\dbar}= c_{\abar\bbar cd}\Lambda_{ab\cbar\dbar}= c_{\abar\bbar\cbar d}\Lambda_{abc\dbar}=0\,.
\eeq
Then, the corresponding beta functions, $\beta_{c_{abcd}\Lambda_{\abar\bbar\cbar\dbar}}$,  $\beta_{ c_{\abar\bbar cd}\Lambda_{ab\cbar\dbar}}$, and  $\beta_{c_{\abar\bbar\cbar d}\Lambda_{abc\dbar}}$ must vanish to all orders of perturbation theory.   Applying this relation to \eq{betaLamabcd}, we note that the symmetry will also impose \textit{separate} independent relations among the scalar potential parameters $\Lambda_{cd\abar\bbar}$, $\Lambda_{\abar\bbar\cbar\dbar}$,and $\Lambda_{d\abar\bbar\ebar}$.  Hence, one may conclude that two separate relations must be satisfied:
\beqa
c_{abcd}(\Lambda_{\abar\bbar\ebar\fbar}\Lambda_{ef\cbar\dbar}+\Lambda_{\abar\cbar\ebar\fbar}\Lambda_{ef\bbar\dbar}+\Lambda_{\abar\dbar\ebar\fbar}\Lambda_{ef\bbar\cbar})\bigl|_{\rm sym}&=& 0\,, \label{sym4}\\
c_{abcd}(\Lambda_{f\abar\bbar\ebar}\Lambda_{e\fbar\cbar\dbar}+\Lambda_{f\abar\cbar\ebar}\Lambda_{e\fbar\bbar\dbar}+\Lambda_{f\abar\dbar\ebar}\Lambda_{e\fbar\bbar\cbar})\bigl|_{\rm sym}&=& 0\,,\label{sym5}
\eeqa
where ``sym'' indicates that the conditions specified by \eq{linearconds} have been imposed on the quartic coupling parameters.
Since we have also imposed CP conservation, it follows that all the quartic couplings in \eqs{sym4}{sym5} are real.
Moreover, since $\Lambda_{\abar\bbar\cbar\dbar}$ and  $\Lambda_{ab\cbar\dbar}$ are independent,
then \eq{sym4} must hold if we numerically set $\Lambda_{ab\cbar\dbar}=\Lambda_{\abar\bbar\cbar\dbar}$ in \eq{sym4}, as justified below \eq{twoloopm2}.
The end result is that \eq{sym4}, where all quartic couplings are real with $\Lambda_{ab\cbar\dbar}=\Lambda_{\abar\bbar\cbar\dbar}$, is algebraically identical to \eq{betalambda}.  Thus, it follows that
\beq
\beta^{I}_{c_{ijk\ell}\lambda_{ijk\ell}}\bigl|_{\rm sym}=c_{ijk\ell}(\lambda_{ijmn}\lambda_{mnk\ell}+\lambda_{ikmn}\lambda_{mnj\ell}+\lambda_{i\ell mn}\lambda_{mnjk})\bigl|_{\rm sym}=0\,,
\eeq
with an implicit sum over repeated indices $i,j,k,\ell\in\{1,2\}$, with $c_{2222}=c_{1111}$, $c_{1222}=-c_{1112}$ and all other $c_{ijk\ell}$ equal to zero.
That is, the one-loop beta function relation satisfied by the scalar potential given in \eq{eq:pott} is explained by the symmetries of the complexified theory given by \eq{V2plusV4}.

Finally, we consider the corresponding two-loop beta function for the $\lambda_{ijk\ell}$ given in \eq{twoloopbetalambda}.
Applying this result to the original model of two real scalar fields subject to the conditions specified in \eq{realconds}, we see that the following two conditions separately hold:
\beqa
\sum_{\rm perm}\lambda_{inpq}\lambda_{mnpq}\lambda_{mjk\ell}\bigl|_{\rm sym}&=&0\,,\label{twolooplam1} \\
\sum_{\rm perm}\lambda_{ijmn}\lambda_{kmpq}\lambda_{\ell npq}\bigl|_{\rm sym}&=&0\,.\label{twolooplam2}
\eeqa
Using the same procedure as before, we can identify two relations (among many) that are the consequence of the vanishing of the two-loop beta function of the complexified theory,
\beq
c_{abcd}\sum_{\rm perm}\Lambda_{\abar\fbar\gbar\hb}\Lambda_{efgh}\Lambda_{\ebar\bbar\cbar\dbar}\bigl|_{\rm sym}&=&0\,,\label{twoloopLamc1} \\
c_{abcd}\sum_{\rm perm}\bigl(\Lambda_{\abar\bbar\ebar\fbar}\Lambda_{eg\cbar\hb}\Lambda_{fh\gbar\dbar}
+\kappa\Lambda_{\abar\bbar ef}\Lambda_{gh\cbar\ebar}\Lambda_{\dbar\fbar\gbar\hb}\bigr)\bigl|_{\rm sym}&=&0\,,\label{twoloopLamc2}
\eeqa
where $\kappa$ is a number that can be evaluated explicitly by expressing the two complex scalars of the complexified theory in terms of the four real fields defined in \eq{complexreal} and then making use of \eq{twoloopbetalambda}.  However, we do not need to know an explicit value for $\kappa$.
We again follow the procedure outlined below \eq{twoloopm2}
where we take all quartic couplings real and numerically set $\Lambda_{ab\cbar\dbar}=\Lambda_{\abar\bbar\cbar\dbar}$ in \eq{twoloopLamc2}.
It then follows that \eqs{twolooplam1}{twoloopLamc1} are algebraically equivalent.  Likewise  \eqs{twolooplam2}{twoloopLamc2}  differ only by an overall numerical factor, which
is irrelevant as both expressions are equal to zero.

\section{Complexification and realification revisited}
\label{sec:revis}

It is perhaps useful to comment on the use of the terms ``complexification'' and ``realification'' used in this paper.   Here, we are employing these terms by analogy with the way they are
used in mathematics.  In particular, these concepts are of particular importance in the theory of Lie algebras~\cite{Baeuerle:1990sm,Onishchik}.

We briefly review the complexification and realification of a Lie algebra by providing some simple examples~\cite{Baeuerle:1990sm}.
Consider the real Lie algebra corresponding to the set of real traceless $2\times 2$ matrices, denoted by $\mathfrak{sl}(2,\mathbb{R})$.   Any real traceless $2\times 2$ matrix is a real linear combination of three generators,
$\{\left(\begin{smallmatrix} 0& 1\\ 0 & 0\end{smallmatrix}\right)\,,\left(\begin{smallmatrix} 0& 0\\ 1 &  0\end{smallmatrix}\right)\,,\left(\begin{smallmatrix} 1& \phm 0\\ 0 & -1\end{smallmatrix}\right)\}$.  The complexification of $\mathfrak{sl}(2,\mathbb{R})$ consists of taking \textit{complex} linear combinations of the generators.   This procedure yields $\mathfrak{sl}(2,\mathbb{C})$, the Lie algebra of complex traceless matrices.   Note that the real dimension of the original Lie algebra has been doubled since ${\rm dim}_{\mathbb{R}}\,\mathfrak{sl}(2,\mathbb{R})=3$, whereas
 ${\rm dim}_{\mathbb{R}}\,\mathfrak{sl}(2,\mathbb{C})=6$.

Continuing with our example of $\mathfrak{sl}(2,\mathbb{C})$, consider the process of realification.   What this means is that we now consider $\mathfrak{sl}(2,\mathbb{C})$ as a real Lie algebra, sometimes denoted by $\mathfrak{sl}(2,\mathbb{C})_{\mathbb{R}}$, consisting of arbitrary real linear combinations of six generators, $\{\left(\begin{smallmatrix} 0& 1\\ 0 & 0\end{smallmatrix}\right)\,,\left(\begin{smallmatrix} 0& 0\\ 1 & 0\end{smallmatrix}\right)\,,\left(\begin{smallmatrix} 1& \phm 0\\ 0 & -1\end{smallmatrix}\right),\left(\begin{smallmatrix} 0& i\\ 0& 0\end{smallmatrix}\right)\,,\left(\begin{smallmatrix} 0& 0\\ i & 0\end{smallmatrix}\right)\,,\left(\begin{smallmatrix} i& \phm 0\\ 0 & -i\end{smallmatrix}\right)\}$.  This is a simple rewriting of the original $\mathfrak{sl}(2,\mathbb{C})$ Lie algebra,\footnote{Indeed, $\mathfrak{sl}(2,\mathbb{C})_{\mathbb{R}}$ is isomorphic to the six-dimensional real Lie algebra of the Lorentz group, $\mathfrak{so}(3,1)$, a fact that plays a significant role in relativistic quantum field theory.} 
so ${\rm dim}_{\mathbb{R}}\,\mathfrak{sl}(2,\mathbb{C})_{\mathbb{R}}={\rm dim}_{\mathbb{R}}\,\mathfrak{sl}(2,\mathbb{C})=6$.

 Note that the realification of a complex Lie algebra $\mathfrak{g}$, denoted by $\mathfrak{g}_{\mathbb{R}}$, should not be confused with a \text{real form} of~$\mathfrak{g}$.   The latter is defined as a subalgebra of $\mathfrak{g}_{\mathbb{R}}$  whose complexification is isomorphic to~$\mathfrak{g}$.   In particular, the dimension of a real form of a complex Lie algebra $\mathfrak{g}$ is equal to $\half\dim_{\mathbb{R}}\,\mathfrak{g}$, whereas ${\rm dim}_{\mathbb{R}}\,\mathfrak{g}_{\mathbb{R}}={\rm dim}_{\mathbb{R}}\,\mathfrak{g}$.  For example,
 the three-dimensional real Lie algebra $\mathfrak{sl}(2,\mathbb{R})$ is an example of a real form of $\mathfrak{sl}(2,\mathbb{C})$, which is clearly distinct from
 the six-dimensional real Lie algebra $\mathfrak{sl}(2,\mathbb{C})_{\mathbb{R}}$.  Finally, we note that one can complexify a complex Lie algebra by complexifying its realification.
 For example, in order to complexify $\mathfrak{sl}(2,\mathbb{C})$, one can complexify
 the real Lie algebra $\mathfrak{sl}(2,\mathbb{C})_{\mathbb{R}}$.  The resulting complex Lie algebra is $\mathfrak{sl}(2,\mathbb{C})\oplus\mathfrak{sl}(2,\mathbb{C})\iso \mathfrak{so}(4,\mathbb{C})$, whose dimension is twice that of $\mathfrak{sl}(2,\mathbb{C})$.

When we complexify a theory of $n$ real scalar fields, the corresponding complexified theory is a theory of $n$ complex scalar fields with twice the number of real degrees of freedom, in analogy with the complexification of a real Lie algebra.
The realification of a theory of complex scalars is obtained by writing $\Phi_a=(\varphi_{a1}+i\varphi_{a2})/\sqrt{2}$ and re-expressing the Lagrangian in terms of the real scalars $\varphi_{1a}$ and~$\varphi_{2a}$, for $a=1,2,\ldots,n$.   This is analogous to the realification of a complex Lie algebra discussed above.  In contrast, the original real scalar field theory whose complexification yields the theory of complex scalars $\Phi_a$ is analogous to the real form of a complex Lie algebra.
Note that starting from a complex scalar field theory, one can perform the complexification process in two steps. First, the realification of the initial complex scalar field theory is performed.
One can then complexify the resulting realified model, in analogy with the complexification of a complex Lie algebra $\mathfrak{sl}(2,\mathbb{C})$ mentioned above.

This leads to the following question: starting from a quantum field theory of $n$ complex scalars, how does one produce a quantum field theory of $n$ real scalars, whose complexification yields the original complex scalar field theory?  The toy example of Section~\ref{sec:toy} and its complexification given in Section~\ref{sec:comp} provide an answer.  Suppose one is given a scalar potential of the form exhibited in \eq{eq:VC}.   Construct from this a real scalar field theory using the following recipe.   First, we replace the kinetic energy term [\eq{KE}] with a canonically normalized kinetic energy term of a real scalar field theory of the form $\half\partial_\mu\varphi_i\partial^\mu\varphi_i$.   Next, we retain only the terms of \eq{eq:VC} that are holomorphic in the complex fields.   That is, we retain $M^2_{\abar\bbar} \Phi_a \Phi_b +\Lambda_{\abar\bbar\cbar\dbar}\,\Phi_a\Phi_b\Phi_c\Phi_d$, while discarding all other terms in \eq{eq:VC}.  Finally, replace the $\Phi_a$ with the same number of real scalar fields $\varphi_a$.   The resulting theory is described by a Lagrangian of a real scalar field theory of the form given by \eq{eq:VR}, whose complexification yields the Lagrangian specified in \eqs{KE}{eq:VC}.   As an example, we noted at the end of Section~\ref{sec:comp} the relation between \eq{Vcholo} and the scalar potential of the toy model of two real scalar fields given in \eq{eq:pott}.

The above procedure suggests an algorithm for constructing examples of real scalar field theories with an RG-stable parameter relation without a symmetry to explain its RG-stability.   Start with a theory of $n$ complex scalars with parameter relations whose RG-stability can be accounted for by the symmetries of the model.  From this theory, construct the corresponding theory of $n$ real scalars whose complexification yields the theory of $n$ complex scalars (using the method outlined above).   If the symmetries of the complexified theory involve some symmetry transformation group that cannot be embedded in O($n)$, then such symmetries cannot survive as a legitimate symmetry of the theory of $n$ real scalars.   This is precisely what happened in Section~\ref{sec:comp}, where
the symmetry employed [\eq{eq:sym}] corresponded to the matrix $\left(\begin{smallmatrix} \phm 0 & i \\  -i & 0\end{smallmatrix}\right)$, which is not an element of O(2).
Nevertheless, the vanishing of the beta functions of the parameter relations of the theory of $n$ complex scalars will still ensure the vanishing of the corresponding beta functions of the theory of $n$ real scalars, as discussed in Section~\ref{sec:fixedpoints}.

\section{Summary of Results}
\label{sec:conc}

A symmetry imposed on a Lagrangian yields relations among its parameters, and those relations will be preserved under renormalization.
In particular, the relations among specific parameters will be obeyed by the beta functions of those parameters, to all orders
of perturbation theory. Recently, in the context of two Higgs doublet models, an example was found that
showed how specific relations between 2HDM parameters were preserved to all orders of perturbation theory, but none of the
known six possible global symmetries of the model could reproduce said relations~\cite{Ferreira:2023dke}. These relations were shown to
be preserved to all orders by the scalar and gauge sectors, and at least up to two loops when the Higgs-fermion Yukawa couplings are included.
To the best of our knowledge, this is the first example of how the RG-stability of a model parameter relation to all orders of perturbation theory may not imply the existence
of a symmetry of the Lagrangian. But if not a symmetry, then what could be causing this remarkable behavior? The authors of Ref.~\cite{Ferreira:2023dke}
observed that this result could formally be obtained by transformations among the real scalar components of the two doublets that involve imaginary numbers. Even stranger, these transformations required that the spacetime coordinates 
transformed into themselves multiplied by an imaginary number to preserve the kinetic energy terms of the model Lagrangian. These transformations correspond to no known legitimate symmetry, and a different explanation for the RG stability of the parameter relations of the model is clearly needed.

In this paper we considered a toy model containing two real scalar fields, and observed that
the RG-stability of a relation among the
parameters of the theory exists that is analogous to that of the 2HDM of Ref.~\cite{Ferreira:2023dke}. 
The relations among the quartic couplings could be reproduced (as in Ref.~\cite{Ferreira:2023dke}) by a simple set of parity transformations on the real fields
(all of them contained in the O(2) group of possible field transformations), but the RG-stable relation between the squared-mass parameters,
$m^2_{22} = - m^2_{11}$ [in the notation of eq.~\eqref{eq:VR}]
cannot be obtained by any known symmetry. However, it can be
reproduced by adopting a ``GOOFy" transformation analogous to those of Ref.~\cite{Ferreira:2023dke}, wherein both scalar fields transform
among themselves multiplied by factors of $i$. These transformations, given in eq.~\eqref{eq:goof},
are not legitimate symmetry transformations of real scalar fields, but they served as inspiration for a possible explanation of the RG stability of the squared-mass parameter relation of the model.
Namely, we promoted the two real scalar fields to two complex scalar fields (a process that was called {\em complexification}),
and imposed simple
symmetries on the resulting complexified model.
These symmetries consisted of an overall parity symmetry to eliminate linear and cubic terms in the scalar potential, CP conservation to enforce reality of the scalar potential parameters,
and an exchange symmetry between the two complex fields involving imaginary numbers [see eq.~\eqref{eq:sym}]. The latter is analogous to
the GOOFy transformation of the real scalar fields of the original toy model, but
in the context of the complexified model this is now a legitimate symmetry.

In the complexified model, the symmetries impose relations among its parameters [given by \eqst{symcond1}{symcond4}] that are
preserved under renormalization and thus yield analogous relations for the corresponding beta functions. Indeed, relations of the
form $c_{ab} M^2_{\bar{a}\bar{b}} = 0$ exist for the squared-mass parameters of the complexified model 
due to the presence of a symmetry which then imply
\beq \label{betasym}
\beta_{c_{ab} M^2_{\bar{a}\bar{b}}}\bigl|_{\rm sym} = 0\,,
\eeq
to all orders of perturbation theory when the parameter relations imposed by the symmetry have been employed (as indicated by the subscript ``sym'').   In fact, one can obtain even stronger conditions beyond what appears in \eq{betasym}.  At any fixed order in perturbation theory,
\eq{betasym} takes the following schematic form:
\beq \label{eff}
\beta_{c_{ab} M^2_{\bar{a}\bar{b}}}=c_{ab}\sum_k f_k(M^2,\Lambda)_{\abar\bbar}\,,
\eeq
where the $f_k(M^2,\Lambda)$ are functions of the squared-mass and self coupling parameters.   Each term in the sum will contain one factor of $M^2$ and $n$ factors of $\Lambda$ at order $n$ in the perturbation expansion.  The tensor $M^2$ can have index structure $cd$, $\cbar\dbar$, or $c\dbar$, and the tensor $\Lambda$ can have index structure $cdef$, $cde\fbar$, $cd\ebar\fbar$, $c\dbar\ebar\fbar$, or $\cbar\dbar\ebar\fbar$.  By appropriate choices of the index structure
along with some appropriate Kronecker deltas to tie together some unbarred/barred index pairs,
the index structure of the $f_k$ must be $\abar\bbar$ as indicated in \eq{eff}.  As a simple example, at one-loop order \eq{eff} takes the form
\beq
\beta_{c_{ab} M^2_{\bar{a}\bar{b}}}\bigl|_{\rm sym} =c_{ab}\left[4M_{\cbar\dbar}^2\Lambda_{cd\abar\bbar}+24M_{cd}^2\Lambda_{\abar\bbar\cbar\dbar}+
6M^2_{e\dbar}\Lambda_{d\abar\bbar\ebar}\right]\bigl|_{\rm sym} = 0\,.
\label{eq:oomph}
\eeq
Although the number of possible terms for $f_k$ expands quickly with each order in perturbation theory, the critical observation is that the $f_k$ are linearly independent tensors.   This means that
\beq \label{stronger}
c_{ab} f_k(M^2,\Lambda)_{\abar\bbar}\bigl|_{\rm sym}=0\,,
\eeq
for each $k$ separately.  This is a stronger result than the one given in \eq{betasym}.

The second critical observation is that there will always be at least one special value of $k$ where $f_k(M^2,\Lambda)$ depends on tensors with an even number of unbarred and barred indices, respectively.
For example, at one loop order, \eq{stronger} takes the form
\beq \label{eq:oomph2}
c_{ab}M_{cd}^2\Lambda_{\abar\bbar\cbar\dbar}\bigl|_{\rm sym}=0\,.
\eeq
Moreover, having imposed CP conservation on the complexified theory, tensors with only unbarred indices are equal to the corresponding tensors with only barred indices.
Beyond one loop order,  $f_k(M^2,\Lambda)$ will also involve $\Lambda$ with two unbarred and two barred indices.  However, since \eq{stronger} is satisfied in general, it also must be satisfied in the special case where $\Lambda_{ab\cbar\dbar}$ is set equal to $\Lambda_{abcd}$.
The end result, is that for the special values of $k$ identified above, an equation of the form given by \eq{stronger} holds where all barred indices are replaced by unbarred indices (and the usual implicit sum over unbarred/barred index pairs is carried out).

The structure of the equations for the beta functions of the original toy model of two real scalar fields also involves sums of linearly independent products of tensors.   The observed RG stability of the parameter relation $m_{22}^2=-m_{11}^2$ (which is not the result of a legitimate symmetry of the original toy model) yields equations that are algebraically equivalent to \eq{stronger} for the special values of $k$ noted above.  For example, at one-loop order,
\beq
\beta_{c_{ij} m^2_{ij}}\bigl|_{\rm sym}  = c_{ij} m_{k\ell}^2\lambda_{ijk\ell}\bigl|_{\rm sym}=0\,,
\label{eq:oomph3}
\eeq
which is algebraically equivalent to \eq{eq:oomph2} after dropping the distinction between unbarred and barred indices.
Thus, we have succeeded in explaining the RG stability of
$m_{22}^2=-m_{11}^2$ as being the result of an ``inherited" symmetry that was imposed on the corresponding complexified theory.

\section{Future Directions}
\label{sec:future}

It would be quite useful to obtain further examples of RG stable parameter relations that cannot be explained by a symmetry of the original theory.\footnote{It is interesting to note a similar phenomenon in Ref.~\cite{Houtz:2016jxk} where
relations between running coupling and masses that do not follow from symmetries were engineered by making use of infrared fixed points of gauge couplings.}
 An algorithm for producing such examples was discussed in Section~\ref{sec:revis}.  In particular, it would also be interesting to apply the ideas of this paper to understand the origin of the RG stability of the parameter relation $m_{22}^2=-m_{11}^2$ in the context of the
 2HDM that was discovered in Ref.~\cite{Ferreira:2023dke}.
  Although we expect that the results of this paper can be used in the 2HDM, there are a number of challenges to confront.  First, since the realification of the 2HDM yields a theory of eight
 real scalar fields, the corresponding complexified theory will be a theory of eight complex fields (or equivalently sixteen real fields).
It is not clear exactly how the SU(2)$_L$ doublet structure of 2HDM scalar fields is manifested in the complexified theory.  For example, is the complexified theory equivalent to a four Higgs doublet model?  Indeed, the strange form of \eq{eq:trandou} suggests that the transformations considered might need to ``break" the
doublet structure somehow, before it is put back together.

Second, the all-order RG invariance found in Ref.~\cite{Ferreira:2023dke} involved not
only the scalar sector but also the gauge interactions; fermions were found to respect the RG fixed points up to two loops via an explicit
calculation, strongly suggesting an all-orders RG invariance. The argument presented here pertains to the scalar sector only, and therefore the first step
would be to verify how the process of complexification of a scalar field theory impacts the
scalar couplings to gauge fields and fermion fields.
Since the interaction of the scalars to gauge fields is generated by replacing the derivatives in the scalar kinetic energy terms [\eq{KE}] with gauge covariant derivatives, it appears that the interactions of the gauge bosons with the scalars of the complexified theory can be treated in a straightforward manner.  The introduction of Yukawa interactions in the complexified theory also seems rather straightforward, but this needs to be checked.

In both the 2HDM and now in the toy model of two real scalar fields examined here, the RG stability of the parameter relations of these theories were discovered, and shown to be
valid to all orders in perturbation theory.  In both cases, the RG stability could not be attributed to a legitimate symmetry of the model.
This paper shows how such parameter relations
may be understood as arising from a symmetry present in the complexified version of the theory.
The beta function relations that yield the RG stability of the corresponding parameter relations of the original model
are therefore understood by virtue of the fact that they are algebraically identical
to symmetry-protected relations of the complexified model. This is still a strange state of affairs, and highly counterintuitive. Why should the RG stability of parameter relations of
a given theory be governed by symmetries of a theory containing a larger field content? And yet that is the strong implication of the work presented here.
We look forward to finding additional examples of these non-symmetry-guaranteed, all-orders-protected, RG invariant relations, and find it fascinating
that such a novel approach to symmetries is still possible, even after all the developments of quantum field theory over the last half century.

\section*{Note added in proof}

After this article was accepted for publication,  
another approach to explaining the RG-stability of parameter relations imposed by GOOFy transformations was suggested by Trautner in Ref.~\cite{Trautner:2025yxz}, even though the corresponding GOOFy scalar field transformations are explicitly broken by the gauge-kinetic energy terms.

\section*{Acknowledgments}

We are grateful for a number of useful discussions with Nathaniel Craig, Bohdan Grzadkowski, Igor Ivanov, Odd Magne Ogreid, John Terning, Jesse Thaler, and Andreas Trautner.
Moreover, H.E.H. acknowledges the kind hospitality and support of Jo\~{a}o Silva, the Instituto Superior T\'{e}cnico, Universidade de Lisboa, and the stimulating atmosphere of the 2024 Workshop on Multi-Higgs Models where this work was initiated.
H.E.H. is supported in part by the U.S. Department of Energy Grant
No.~\uppercase{DE-SC}0010107, and in part by grant NSF PHY-2309135 to the Kavli Institute for Theoretical Physics (KITP).  H.E.H. greatly appreciates the support of the KITP where this work was completed.   P.M.F. is supported
by \textit{Funda\c c\~ao para a Ci\^encia e a Tecnologia} (FCT)
through contracts
UIDB/00618/2020, UIDP/00618/2020, CERN/FIS-PAR/0025/2021 and 2024.03328.CERN.

\section*{Additional acknowledgments}

We thank Patrick Fox for pointing out the error in the original published
version of \eq{symcond1} and for other insightful remarks in discussions
with H.E.H. at the Munich Institute for Astro-, Particle and BioPhysics
(MIAPbP), which is funded by the Deutsche Forschungsgemeinschaft
(DFG, German Research Foundation) under Germany's Excellence 
Strategy – EXC-2094–390783311, and at the Aspen Center for Physics,
which is supported by the National Science Foundation Grant
No. PHY-2210452.  
H.E.H also greatly appreciates the hospitality and housing support of
the NYU Center for Cosmology and Particle Physics during two recent
visits, where he benefited from stimulating conversations
with Ken Van Tilburg and his collaborators concerning the use of the
spurion technique in determining RG fixed points.

\begin{appendices}

\section{Beta functions of a toy model of two real scalar fields}
\label{app:beta}

Consider a quantum field theory of two real scalar fields governed by the Lagrangian specified in \eq{eq:VR}.  At two-loop order, the beta functions of the parameters $m^2_{ij}$ and $\lambda_{ijk\ell}$ for
$i, j, k, \ell\in\{1,2\}$ are denoted by $\beta=\beta^{I}+\beta^{II}$, where the corresponding one-loop and two-loop contributions are exhibited in \eqst{betam2}{twoloopbetalambda}.
In this Appendix, we provide the corresponding analytic expressions for the beta functions of $m_{11}^2$, $m_{22}^2$, $\lambda_{1111}$, $\lambda_{2222}$, $\lambda_{1112}$, and $\lambda_{1222}$.    We then demonstrate that in the toy model with
\beq \label{parmrels}
m_{22}^2=-m_{11}^2\,,\qquad\quad \lambda_{1111}=\lambda_{2222}\,,\qquad\quad \lambda_{1112}=-\lambda_{1222}\,,
\eeq
these parameter relations are stable under renormalization group running, since the corresponding beta functions for $m_{11}^2+m_{22}^2$, $\lambda_{1111}-\lambda_{2222}$ and $\lambda_{1112}+\lambda_{1222}$ vanish exactly.

We first evaluate the one-loop beta functions of $m_{11}^2$ and $m_{22}^2$.  Using \eq{betam2},
\beqa
\beta^I_{m_{11}^2}&=&m_{11}^2\lambda_{1111}+m_{22}^2\lambda_{1122}+2m_{12}^2\lambda_{1112}\,, \\
\beta^I_{m_{22}^2}&=&m_{22}^2\lambda_{2222}+m_{11}^2\lambda_{1122}+2m_{12}^2\lambda_{1222}\,.
\eeqa
It then follows that
\beq \label{betamI}
\beta^I_{m_{11}^2+m_{22}^2}=\beta^I_{m_{11}^2}+\beta^I_{m_{22}^2}=m_{11}^2\lambda_{1111}+m_{22}^2\lambda_{2222}+(m_{11}^2+m_{22}^2)\lambda_{1122}+2m_{12}^2(\lambda_{1112}+\lambda_{1222})\,.\phantom{xxxx}
\eeq
After imposing the parameter relations of \eq{parmrels} [denoted below by the subscript ``sym''], we obtain
\beq
\beta^I_{m_{11}^2+m_{22}^2}\bigl|_{\rm sym}=0\,.
\eeq

Next, we evaluate the one-loop beta functions of $\lambda_{1111}$, $\lambda_{2222}$, $\lambda_{1112}$, and $\lambda_{2222}$.  Using \eq{betalambda},
\beqa
\beta^I_{\lambda_{1111}}&=& 3(\lambda_{1111}^2+2\lambda_{1112}^2+\lambda_{1122}^2)\,, \\
\beta^I_{\lambda_{2222}}&=& 3(\lambda_{2222}^2+2\lambda_{1222}^2+\lambda_{1122}^2)\,, \\
\beta^I_{\lambda_{1112}}&=& 3(\lambda_{1111}\lambda_{1112}+2\lambda_{1112}\lambda_{1122}+\lambda_{1122}\lambda_{1222})\,, \\
\beta^I_{\lambda_{1222}}&=& 3(\lambda_{1112}\lambda_{1122}+2\lambda_{1122}\lambda_{1222}+\lambda_{1222}\lambda_{2222})\,.
\eeqa
It follows that
\beqa
\beta^I_{\lambda_{1111}-\lambda_{2222}}&=&\beta^I_{\lambda_{1111}}-\beta^I_{\lambda_{2222}}=3\bigl[\lambda_{1111}^2-\lambda_{2222}^2+2\bigl(\lambda_{1112}^2-\lambda_{1222}^2\bigr)\bigr]\,,\label{betalamI}\\
\beta^I_{\lambda_{1112}+\lambda_{1222}}&=&\beta^I_{\lambda_{1112}}+\beta^I_{\lambda_{1222}}=3\bigl[\lambda_{1111}\lambda_{1112}+\lambda_{2222}\lambda_{1222}+3\lambda_{1122}(\lambda_{1112}+\lambda_{1222})\bigr]\,.\label{betalam2I}
\eeqa
After imposing the parameter relations of \eq{parmrels}, we obtain
\beq
\beta^I_{\lambda_{1111}-\lambda_{2222}}\bigl|_{\rm sym}=\beta^I_{\lambda_{1112}+\lambda_{1222}}\bigl|_{\rm sym}=0\,.
\eeq

To compute the two-loop contributions to the beta functions of $m_{ij}^2$ and $\lambda_{ijk\ell}$ given in \eqs{twoloopbetam}{twoloopbetalambda}, we shall evaluate the following quantities:
\beqa
A_{ij}& \equiv & \half\bigl[\lambda_{ik\ell m}\lambda_{nk\ell m}m^2_{nj}+\lambda_{jk\ell m}\lambda_{nk\ell m}m^2_{ni}\bigr]\,, \\
B_{ij}& \equiv & m^2_{k\ell}\lambda_{ikmn}\lambda_{j\ell mn}\,, \\
C_{ijk\ell}& \equiv &\frac{1}{24} \sum_{\rm perm}\lambda_{inpq}\lambda_{mnpq}\lambda_{mjk\ell}\,, \\
D_{ijk\ell}& \equiv &\frac{1}{24} \sum_{\rm perm}\lambda_{ijmn}\lambda_{kmpq}\lambda_{\ell npq}\,,
\eeqa
where ``perm'' indicates that the sum includes terms in which the uncontracted indices $i$, $j$, $k$, and~$\ell$ have been permuted in all possible ways.  In addition, there are implicit sums over each repeated index pair.  For example, in light of \eq{twoloopbetam},
\beq \label{betaIIm11mww}
\beta^{II}_{m_{11}^2+m_{22}^2}=\tfrac16\bigl(A_{11}+A_{22}\bigr)-2\bigl(B_{11}+B_{22}\bigr)\,.
\eeq

We then obtain
\beqa
A_{11}&=&\bigl[\lam_{1111}^2+3\bigl(\lam_{1112}^2+\lam_{1122}^2\bigr)+\lam_{1222}^2\bigr]m_{11}^2+\bigl[\lam_{1112}\bigl(\lam_{1111}+3\lam_{1122}\bigr)+\lam_{1222}\bigl(\lam_{2222}+3\lam_{1122}\bigr)\bigr]m_{12}^2\,,\nn  \\
&& \phantom{line}\label{a11} \\
A_{22}&=&\bigl[\lam_{1112}^2+3\bigl(\lam_{1122}^2+\lam_{1222}^2\bigr)+\lam_{2222}^2\bigr]m_{22}^2+\bigl[\lam_{1112}\bigl(\lam_{1111}+3\lam_{1122}\bigr)+\lam_{1222}\bigl(\lam_{2222}+3\lam_{1122}\bigr)\bigr]m_{12}^2\,,\nn \\
&& \phantom{line} \label{a22} \\
B_{11}&=&\bigl(\lam_{1111}^2+2\lam_{1112}^2+\lam_{1122}^2\bigr)m_{11}^2+\bigl(\lam_{1112}^2+2\lam_{1122}^2+\lam^2_{1222}\bigr)m^2_{22}
\nn \\
&& \qquad\qquad +2\bigl(\lam_{1111}\lam_{1112}+2\lam_{1112}\lam_{1122}+\lam_{1122}\lam_{1222}\bigr)m_{12}^2\,,\label{b11} \\
B_{22}&=& \bigl(\lam_{1112}^2+2\lam_{1122}^2+\lam_{1222}^2\bigr)m_{11}^2+\bigl(\lam_{1122}^2+2\lam_{1222}^2+\lam_{2222}^2\bigr)m^2_{22}
\nn \\
&& \qquad\qquad +2\bigl(\lam_{1112}\lam_{1122}+2\lam_{1122}\lam_{1222}+\lam_{1222}\lam_{2222}\bigr)m_{12}^2\,.\label{b22}
\eeqa
One can see by inspection that
\beq \label{aabb}
A_{11}+A_{22}\bigl|_{\rm sym}=0\,,\qquad\quad B_{11}+B_{22}\bigl|_{\rm sym}=0\,,
\eeq
where ``sym'' again indicates that the parameter relations given by \eq{parmrels} have been employed.
These results confirm the vanishing of the two-loop contribution to the beta function of $m_{11}^2+m_{22}^2$.

It is instructive to use the above results to compute the beta function of the parameter $m_1^2$ that appears in \eq{complexLag}. In light of \eq{rel1}, it follows that
$\beta_{m_1^2}=\half\beta_{m_{11}^2+m_{22}^2}$.  We now invert the results of \eqst{rel1}{rel5} to obtain
\beq
m_{11}^2=m_1^2+2\Re m_2^2\,,\qquad \quad  m_{22}^2=m_1^2-2\Re m_2^2\,,\qquad \quad m^2_{12}=-2\Im m_2^2\,,
\eeq
and
\beqa
\lambda_{1111} &=& 6\lambda_1+12\Re(\lambda_2+\lambda_3)\,,\\
\lambda_{1112} &=& -12\Im\lambda_2-6\Im\lambda_3\,,\\
\lambda_{1122} &=& 2\lambda_1-12\lambda_2\,, \\
\lambda_{1222} &=& 12\Im\lambda_2-6\Im\lambda_3\,,\\
\lambda_{2222} &=& 6\lambda_1+12\Re(\lambda_2-\lambda_3)\,.
\eeqa
Using eqs.~(\ref{betamI}) and (\ref{betaIIm11mww})--(\ref{b22}), one can obtain the one-loop and two-loop contributions to $\beta_{m_1^2}$,
\beqa
\beta^I_{m_1^2}&=& 8\lambda_1 m_1^2+24\Re(\lambda_3^* m_2^2)\,,\label{betam1one}\\
\beta^{II}_{m_1^2}&=& -8m_1^2\bigl(11\lambda_1^2+132|\lambda_2|^2+33|\lambda_3|^2\bigr)-528\lambda_1\Re(\lambda_3^* m_2^2)-1056\Re(\lambda_2^*\lambda_3 m_2^2)\,,\label{betam1two}
\eeqa
after making use of \eq{rel1}.
Note that \eq{betam1one} can also be derived directly from \eq{betaMsq2}.
As expected, $\beta^I_{m_1^2}=\beta^{II}_{m_1^2}=0$ after imposing the relations $m_1^2=\lambda_3=0$ as indicated in \eq{betazero1}.
\clearpage

The structure of the expressions on the right-hand sides of \eqs{betam1one}{betam1two} is noteworthy.  Using the unbarred/barred index notation introduced in Section~\ref{sec:comp},
one can write
\beqa
m_1^2&\equiv& M^2_{1\bar{1}}\,,\qquad m_2^2\equiv  M^2_{\bar{1}\bar{1}}\,,\qquad   (m_2^2)^*\equiv  M^2_{11}\,, \label{require1}\\
\lambda_1 &\equiv& \Lambda_{11\bar{1}\bar{1}}\,,\qquad \lambda_{2}\equiv \Lambda_{\bar{1}\bar{1}\bar{1}\bar{1}}\,,\qquad \,\,\,\lambda_2^*\equiv \Lambda_{1111}\,,\qquad
\lambda_3\equiv\Lambda_{1\bar{1}\bar{1}\bar{1}}\,,\qquad \lambda_3^*\equiv \Lambda_{111\bar{1}}\,.\label{require2}
\eeqa
In light of the covariance properties of the corresponding beta functions noted at the end of Section~\ref{sec:beta}, it follows that the index structure of $\beta_{m_1^2}$ must contain an equal number of unbarred and barred indices.  One can check that the expressions on the right-hand sides of \eqs{betam1one}{betam1two} satisfy this requirement.  Indeed, this requirement is equivalent to
the result of the spurion analysis of
Ref.~\cite{spurion}.\footnote{We thank Ken Van Tilburg for communicating the results of Ref.~\cite{spurion} to us prior to its release.}  Such an analysis fixes the forms that can
appear in the computation of $\beta_{m_1^2}$ at any loop order, and implies that \eq{betazero1} is satisfied to all orders in perturbation theory.

Next, we note that in light of \eq{twoloopbetalambda},
\beqa
\beta^{II}_{\lambda_{1111}-\lambda_{2222}}&=&\tfrac13\bigl(C_{1111}-C_{2222}\bigr)-6\bigl(D_{1111}-D_{2222})\,, \label{morebetas}\\
\beta^{II}_{\lambda_{1112}+\lambda_{1222}}&=&\tfrac13\bigl(C_{1112}+C_{1222}\bigr)-6\bigl(D_{1112}+D_{1222})\,.
\eeqa
Thus, we record the following results:
\beqa
&& \hspace{-0.5in} C_{1111}=\lam_{1111}^3+\lam_{1111}\bigl(4\lam_{1112}^2+3\lam_{1122}^2+\lam_{1222}^2\bigr)+3\lam_{1112}\lam_{1122}(\lam_{1112}+\lam_{1222})+\lam_{1112}\lam_{1222}\lam_{2222}\,, \label{c1111} \\
&&  \hspace{-0.5in} C_{2222}=\lam_{2222}^3+\lam_{2222}\bigl(4\lam_{1222}^2+3\lam_{1122}^2+\lam_{1112}^2\bigr)+3\lam_{1222}\lam_{1122}(\lam_{1112}+\lam_{1222})+\lam_{1111}\lam_{1112}\lam_{1222}\,, \label{c2222} \\
&&  \hspace{-0.5in} D_{1111}=\lam_{1111}^3+\lam_{1111}\bigl(4\lam_{1112}^2+\lam_{1122}^2\bigr)+2\lam_{1122}^3+\lam_{1122}\bigl(\lam_{1222}^2+2\lam_{1112}\lam_{1222}+5\lam_{1112}^2\bigr)\,, \label{d1111}\\
&&  \hspace{-0.5in} D_{2222}=\lam_{2222}^3+\lam_{2222}\bigl(4\lam_{1222}^2+\lam_{1122}^2\bigr)+2\lam_{1122}^3+\lam_{1122}\bigl(\lam_{1112}^2+2\lam_{1112}\lam_{1222}+5\lam_{1222}^2\bigr)\,.\label{d2222}
\eeqa
After imposing the parameter relations of \eq{parmrels}, one can see by inspection that
\beq \label{ccdd1}
C_{1111}-C_{2222}\bigl|_{\rm sym}=0\,,\qquad\quad D_{1111}-D_{2222}\bigl|_{\rm sym}=0\,.
\eeqa
These results confirm the vanishing of the two-loop contribution to the beta function of $\lambda_{1111}-\lambda_{2222}$.

Finally, we have evaluated the following quantities:
\begingroup
\allowdisplaybreaks
\beqa
C_{1112}&=& \frac{1}{4}\Bigl\{10\lam_{1112}^3+\lam_{1112}\bigl(4\lam_{1111}^2+6\lam_{1111}\lam_{1122}+21\lam_{1122}^2+6\lam_{1222}^2+\lam_{2222}^2\bigr) \nn \\
&& \qquad\qquad +\lambda_{1222}\bigl[3\lambda_{1122}\bigl(\lambda_{1111}+3\lambda_{1122}+\lambda_{2222}\bigr)+\lambda_{1111}\lambda_{2222}\bigr]\Bigr\}\,,\\
C_{1222}&=& \frac{1}{4}\Bigl\{10\lam_{1222}^3+\lam_{1222}\bigl(4\lam_{2222}^2+6\lam_{2222}\lam_{1122}+21\lam_{1122}^2+6\lam_{1112}^2+\lam_{1111}^2\bigr) \nn \\
&& \qquad\qquad +\lambda_{1112}\bigl[3\lambda_{1122}\bigl(\lambda_{2222}+3\lambda_{1122}+\lambda_{1111}\bigr)+\lambda_{1111}\lambda_{2222}\bigr]\Bigr\}\,,\\
D_{1112}&=&  \frac{1}{2}\Bigl\{3\lam_{1112}^3+\lam_{1222}^3+3\lam_{1112}^2\lam_{1222}+\lam_{1112}\bigl(2\lam_{1111}^2+8\lam_{1122}^2+\lam_{1222}^2+\lam_{1122}\lam_{2222}+5\lam_{1111}\lam_{1122}\bigr)
\nn \\
&& \qquad\qquad  +\lam_{1222}\bigl[6\lam_{1122}^2+\lam_{1122}\bigl(\lam_{1111}+\lam_{2222}\bigr)\bigr]\Bigr\}\,,
\\
D_{1222}&=& 
\frac{1}{2}\Bigl\{3\lam_{1222}^3+\lam_{1112}^3+3\lam_{1222}^2\lam_{1112}+\lam_{1222}\bigl(2\lam_{2222}^2+8\lam_{1122}^2+\lam_{1112}^2+\lam_{1122}\lam_{1111}+5\lam_{2222}\lam_{1122}\bigr)
\nn \\
&& \qquad\qquad  +\lam_{1112}\bigl[6\lam_{1122}^2+\lam_{1122}\bigl(\lam_{1111}+\lam_{2222}\bigr)\bigr]\Bigr\}\,. \label{D1222}
\eeqa
\endgroup
After imposing the parameter relations of \eq{parmrels}, one can see by inspection that
\beq \label{ccdd2}
C_{1112}+C_{1222}\bigl|_{\rm sym}=0\,,\qquad\quad D_{1112}+D_{1222}\bigl|_{\rm sym}=0\,.
\eeqa
These results confirm the vanishing of the two-loop contribution to the beta function of $\lambda_{1112}+\lambda_{1222}$.

The results obtained in \eqss{aabb}{ccdd1}{ccdd2} indicate that
the two-loop contributions to the beta functions of $m_{11}^2+m_{22}^2$, $\lambda_{1111}-\lambda_{2222}$, and $\lambda_{1112}+\lambda_{1222}$, respectively, consist of the sum of
two linearly-independent contributions given by \eqst{twoloopbetam}{twoloopbetalambda}, each of which has been shown in this Appendix to separately vanish.

Again, it is instructive to use the above results to compute the beta function of the parameter~$\lambda_3$ that appears in \eq{complexLag}. In light of \eq{rel5}, it follows that
\beq
\beta_{\lambda_3}=\tfrac{1}{24}\beta_{\lambda_{1111}-\lambda_{2222}}-\tfrac{i}{12}\beta_{\lambda_{1112}+\lambda_{1222}}\,.
\eeq
After using eqs.~(\ref{betalamI}), (\ref{betalam2I}), and (\ref{morebetas})--(\ref{D1222}), one can obtain the one-loop and two-loop contributions to $\beta_{\lambda_3}$,
\beqa
\beta^I_{\lambda_3}&=&36\bigl(\lambda_1\lambda_3+2\lambda_2\lambda_3^*\bigr)\,,\label{betafun3I} \\
\beta^{II}_{\lambda_3}&=&-8\bigl[\lambda_3\bigl(79\lambda_1^2+396|\lambda_2|^2+102|\lambda_3|^2\bigr)+276\lambda_1\lambda_2\lambda_3^*\bigr]\,.\label{betafun3II}
\eeqa
Note that \eq{betafun3I} can also be derived directly from \eq{betaLamabcd3}.
As expected, $\beta^I_{\lambda_3}=\beta^{II}_{\lambda_3}=0$ after imposing the parameter relation $\lambda_3=0$ as indicated in \eq{betazero2}.

One can again use the covariance properties of the beta functions noted at the end of Section~\ref{sec:beta} to conclude that, in light of \eqs{require1}{require2},  the index structure of $\beta_{\lambda_3}$ must satisfy the condition that the number of barred indices minus the number of unbarred indices is equal to two.   Indeed, the results of \eqs{betafun3I}{betafun3II} satisfy this requirement, which is again 
consistent with the spurion analysis of Ref.~\cite{spurion}.  Such an analysis fixes the forms that can
appear in the computation of $\beta_{\lambda_3}$ at any loop order, and implies that \eq{betazero2} is satisfied to all orders in perturbation theory.

\section{Stability of parameter relations under a change of the scalar field basis}
\label{app:basis}

Consider the most general renormalizable theory of two real scalar fields with scalar potential
\beqa
V&=&\half m_{11}^2\varphi_1^2+ \half m_{22}^2\varphi_2^2+m_{12}^2\varphi_1\varphi_2 \nonumber \\
&& +\tfrac{1}{24}\lambda_{1111}\varphi_1^4+\tfrac{1}{24}\lambda_{2222}\varphi_2^4+\tfrac14\lambda_{1122}\varphi_1^2\varphi_2^2+\tfrac16\lambda_{1112}\varphi_1^3\varphi_2+\tfrac16\lambda_{1222}\varphi_1\varphi_2^3\,. \label{scalarLag}
\eeq
A scalar field basis transformation is a linear redefinition of the scalar fields that preserves the scalar field kinetic energy terms, 
\beq
\mathscr{L}_{\rm KE}=\half\partial_\mu\varphi_1\partial^\mu\varphi_1+\half\partial_\mu\varphi_2\partial^\mu\varphi_2\,.
\eeq
That is, the most general change of the scalar field basis is an O(2) transformation,
\beq \label{bchange}
\begin{pmatrix} \varphi_1 \\ \varphi_2\end{pmatrix} =\begin{pmatrix} \phm c_\theta & \phm s_\theta \\ -\varepsilon s_\theta & \phm\varepsilon c_\theta\end{pmatrix}\begin{pmatrix} \varphi_1^\prime \\ \varphi_2^\prime\end{pmatrix}\,,
\eeq
where $c_\theta\equiv\cos\theta$,  $s_\theta\equiv\sin\theta$, and the parameter $\varepsilon$ is either $+1$ or $-1$.

Inserting \eq{bchange} into \eq{scalarLag}, we obtain the scalar potential in terms of the primed fields with primed coefficients given by:
\begingroup
\allowdisplaybreaks
\beqa
m^{\prime\,2}_{11} &=& m_{11}^2 c^2_\theta +m_{22}^2 s^2_\theta-\varepsilon m_{12}^2\sin 2\theta\,, \label{e4} \\
m^{\prime\,2}_{22} &=& m_{11}^2 s^2_\theta +m_{22}^2 c^2_\theta+\varepsilon m_{12}^2\sin 2\theta\,, \\
m^{\prime,2}_{12}&=&\half(m_{11}^2-m_{22}^2)\sin 2\theta+\varepsilon m_{12}^2\cos 2\theta\,,\\
\lambda^\prime_{1111} &=& c^4_\theta \lambda_{1111}+s_\theta^4\lambda_{2222}+6s^2_\theta c^2_\theta\lambda_{1122}-4\varepsilon s_\theta c_\theta(c^2_\theta\lambda_{1112}+s_\theta^2\lambda_{1222})\,, \\
\lambda^\prime_{1112} &=& c_\theta s_\theta\bigl(c^2_\theta \lambda_{1111}-s_\theta^2\lambda_{2222}\bigr)-3s_\theta c_\theta(c_\theta^2-s_\theta^2)\lambda_{1122}+\varepsilon c^2_\theta(c^2_\theta-3s_\theta^2)\lambda_{1112}-\varepsilon s^2_\theta(s^2_\theta-3 c^2_\theta)\lambda_{1222},\phantom{xxxx} \\
\lambda^\prime_{1122} &=& c^2_\theta s^2_\theta(\lambda_{1111}+\lambda_{2222})+ (1+2c^2_\theta s^2_\theta)\lambda_{1122}+2\varepsilon s_\theta c_\theta(c^2_\theta-s^2_\theta)\bigl(\lambda_{1112}-\lambda_{1222}\bigr)\,,\\
\lambda^\prime_{1222} &=& c_\theta s_\theta\bigl(s^2_\theta \lambda_{1111}-c_\theta^2\lambda_{2222}\bigr)+3s_\theta c_\theta(c_\theta^2-s_\theta^2)\lambda_{1122}-\varepsilon s^2_\theta(s^2_\theta-3c_\theta^2)\lambda_{1112}+\varepsilon c^2_\theta(c^2_\theta-3 s^2_\theta)\lambda_{1222},\phantom{xxxx} \\
\lambda^\prime_{2222} &=& s^4_\theta \lambda_{1111}+c_\theta^4\lambda_{2222}+6s^2_\theta c^2_\theta\lambda_{1122}+4\varepsilon s_\theta c_\theta(s^2_\theta\lambda_{1112}+c_\theta^2\lambda_{1222})\,.\label{e12}
\eeqa
\endgroup

Consider the parameter relations given in the $\{\Phi_1$\,,\,$\Phi_2\}$ basis by \eq{realconds}, which we repeat below for the benefit of the reader:
\beq \label{app:realconds}
m_{22}^2 =-m_{11}^2\,\qquad\quad\lambda_{1111}=\lambda_{2222}\,,\qquad\quad
\lambda_{1112}= -\lambda_{1222}\,.
\eeq
Plugging these relations into \eqst{e4}{e12} yields the corresponding parameter relations in the $\{\Phi^\prime_1$\,,\,$\Phi^\prime_2\}$ basis:
$m_{22}^{\prime\,2} =-m_{11}^{\prime\,2}$, $\lambda^\prime_{1111}=\lambda^\prime_{2222}$, and
$\lambda^\prime_{1112}=-\lambda^\prime_{1222}$.
That is, the parameter relations in \eq{app:realconds} are RG stable \textit{and} stable under a change of scalar field basis.  In contrast, $m^{\prime\,2}_{12}\neq m^2_{12}$ (assuming a nontrivial change of basis).   This means that one is free to choose $\theta$ such that $m^{\prime\,2}_{12}=0$ at tree level, which corresponds to a choice of $\theta$ such that $\tan 2\theta=\varepsilon m_{12}^2/m_{22}^2$
after making use of the squared mass relation in \eq{app:realconds}.

However, it is noteworthy that the choice of basis needed to set $m^{\prime\,2}_{12}=0$ is not stable under RG running.   In particular, in light of \eq{betam2},
\beq
\beta^I_{m_{12}^2} = m_{11}^2\lambda_{1112}+m_{22}^2\lambda_{1222}+2m_{12}^2\lambda_{1122}\,.
\eeq
After imposing the parameter relations of \eq{app:realconds}, we obtain
\beq
\beta^I_{m_{12}^2}\Bigr|_{\rm sym}=2\bigl(m_{11}^2\lambda_{1112}+m_{12}^2\lambda_{1122}\bigr)\,.
\eeq
That is, if we set $m_{12}^2=0$ at some energy scale $\mu_1$ then $m_{12}^2\neq 0$ at energy scale $\mu_2\neq \mu_1$, which means that the choice of the scalar field basis is not stable with respect to RG running.

The above results should be contrasted with the parameter relations given by \eq{altrealconds},
\beq \label{app:altrealconds}
m^2_{22} =m^2_{11}\,\qquad\quad m_{12}^2=0\,,\qquad\quad \lambda_{1111}=\lambda_{2222}\,,\qquad\quad
\lambda_{1112}=-\lambda_{1222}\,,
\eeq
which are enforced by a legitimate symmetry.
Inserting these relations into \eqst{e4}{e12} yields $m^{\prime\,2}_{22} =m^{\prime\,2}_{11}$, $m_{12}^{\prime\,2}=0$, $\lambda^\prime_{1111}=\lambda^\prime_{2222}$, and
$\lambda^\prime_{1112}=-\lambda^\prime_{1222}$.  That is, the parameter relations in \eq{app:altrealconds}, including the condition $m^2_{12}=0$, are RG stable and
stable under a change of scalar field basis.  In this case, the fact that the choice of scalar field basis is not stable with respect to RG running has no impact on the symmetry-imposed parameter relations.

\section{Parameters of the complexified theory}
\label{app1}

The independent squared-mass and quartic coupling parameters of the complexified theory are listed in \eqs{ten}{thirtyfive} respectively.  If we now express the complex fields $\Phi_1$ and $\Phi_2$ in terms of four real fields $\varphi_i$ defined in \eq{complexreal}, then the scalar potential given by $V_C$ in \eq{eq:VC} can be rewritten as a scalar potential of the corresponding realified theory,
\beq
V_C \,=\, \half m^2_{ij}\,\varphi_i\varphi_j\,+\frac{1}{4!}\,\lambda_{ijk\ell}\,\varphi_i\varphi_j \varphi_k\varphi_\ell\,,\qquad \text{$i,j,k,\ell\in\{1,2,3,4\}$},
\eeq
with an implicit sum over repeated indices.   In particular, $m^2_{ij}$ and $\lambda_{ijk\ell}$ are completely symmetric real tensors, with 10 and 35 independent components, respectively.\footnote{In general, the number of independent components of a completely symmetric real rank $r$ tensor whose indices take on the values $1,2,\ldots,d$ is equal to $(d+r-1)!/[(d-1)! r!]$.}

It is straightforward to express the independent elements of $m^2_{ij}$ in terms of the 10 squared-mass parameters exhibited in \eq{ten}:
\beqa
m_{11}^2&=&M^2_{1\bar{1}}+2\Re M^2_{11}\,,\\
m_{22}^2&=&M^2_{1\bar{1}}-2\Re M^2_{11}\,,\\
m_{33}^2&=&M^2_{2\bar{2}}+2\Re M^2_{22}\,,\\
m_{44}^2&=&M^2_{2\bar{2}}-2\Re M^2_{22}\,,\\
m_{12}^2&=&2\Im M_{11}^2\,,\\
m_{34}^2&=&2\Im M_{22}^2\,,\\
m_{13}^2&=&\Re M^2_{1\bar{2}}+2\Re M^2_{12}\,,\\
m_{24}^2&=&\Re M^2_{1\bar{2}}-2\Re M^2_{12}\,,\\
m_{14}^2&=& -\Im M^2_{1\bar{2}}+2\Im M_{12}^2\,,\\
m_{23}^2&=& \Im M^2_{1\bar{2}}+2\Im M_{12}^2\,.
\eeqa

Likewise, it is straightforward to express the independent elements of $\lambda_{ijk\ell}$ in terms of the 35 self-coupling parameters exhibited in \eq{thirtyfive}:
\begingroup
\allowdisplaybreaks
\beqa
\lam_{1111}&=&6\Lam_{11\bar{1}\bar{1}}+12\Re\bigl(\Lam_{1111}+\Lam_{111\bar{1}}\bigr)\,,
\\
\lam_{1112}&=&6\Im\bigl(2\Lam_{1111}+\Lam_{111\bar{1}}\bigr)\,,
\\
\lam_{1113}&=&3\Re\bigl[2\Lam_{11\bar{1}\bar{2}}+4\Lam_{1112}+\Lam_{111\bar{2}}+3\Lam_{112\bar{1}}\bigr]\,,
\\
\lam_{1114}&=&-3\Im\bigl[2\Lam_{11\bar{1}\bar{2}}-4\Lam_{1112}+\Lam_{111\bar{2}}-3\Lam_{112\bar{1}}\bigr]\,,
\\
\lam_{1122}&=&2\Lam_{11\bar{1}\bar{1}}-12\Re\Lam_{1111}\,,
\\
\lam_{1123}&=&\Im\bigl(2\Lam_{11\bar{1}\bar{2}}+12\Lam_{1112}+3\Lam_{111\bar{2}}+3\Lam_{112\bar{1}}\bigr)\,,
\\
\lam_{1124}&=&\Re\bigl[2\Lam_{11\bar{1}\bar{2}}-12\Lam_{1112}+3\Lam_{111\bar{2}}-3\Lam_{112\bar{1}}\bigr]\,,
\\
\lam_{1133}&=&4\Lam_{12\bar{1}\bar{2}}+2\Re\bigl[\Lam_{11\bar{2}\bar{2}}+6\Lam_{1122}+3\Lam_{112\bar{2}}+3\Lam_{122\bar{1}}\bigr]\,,
\\
\lam_{1134}&=&-2\Im\bigl(\Lam_{11\bar{2}\bar{2}}-6\Lam_{1122}-3\Lam_{122\bar{1}}\bigr)\,,
\\
\lam_{1144}&=&4\Lam_{12\bar{1}\bar{2}}-2\Re\bigl[\Lam_{11\bar{2}\bar{2}}+6\Lam_{1122}-3\Lam_{112\bar{2}}+3\Lam_{122\bar{1}}\bigr]\,,
\\
\lam_{1222}&=&-6\Im\bigl(2\Lam_{1111}-\Lam_{111\bar{1}}\bigr)\,,
\\
\lam_{1223}&=&\Re\bigl[2\Lam_{11\bar{1}\bar{2}}-12\Lam_{1112}-3\Lam_{111\bar{2}}+3\Lam_{112\bar{1}}\bigr]\,,
\\
\lam_{1224}&=&-\Im\bigl[2\Lam_{11\bar{1}\bar{2}}+12\Lam_{1112}-3\Lam_{111\bar{2}}-3\Lam_{112\bar{1}}\bigr]\,,
\\
\lam_{1233}&=& 2\Im\bigl[\Lam_{11\bar{2}\bar{2}}+6\Lam_{1122}+3\Lam_{112\bar{2}}\bigr]\,,
\\
\lam_{1234}&=&2\Re\bigl(\Lam_{11\bar{2}\bar{2}}-6\Lam_{1122}\bigr)\,,
\\
\lam_{1244}&=&-2\Im\bigl(\Lam_{11\bar{2}\bar{2}}+6\Lam_{1122}-3\Lam_{112\bar{2}}\bigr)\,,
\\
\lam_{1333}&=&3\Re\bigl[2\Lam_{12\bar{2}\bar{2}}+4\Lam_{1222}+3\Lam_{122\bar{2}}+\Lam_{222\bar{1}}\bigr]\,,
\\
\lam_{1334}&=& -\Im\bigl[2\Lam_{12\bar{2}\bar{2}}-12\Lam_{1222}-3\Lam_{122\bar{2}}-3\Lam_{222\bar{1}}\bigr]\,,
\\
\lam_{1344}&=&\Re\bigl[2\Lam_{12\bar{2}\bar{2}}-12\Lam_{1222}+3\Lam_{122\bar{2}}-3\Lam_{222\bar{1}}\bigr]\,,
\\
\lam_{1444}&=&-3\Im\bigl[2\Lam_{12\bar{2}\bar{2}}+4\Lam_{1222}-3\Lam_{122\bar{2}}+\Lam_{222\bar{1}}\bigr]\,,
\\
\lam_{2222}&=&6\Lam_{11\bar{1}\bar{1}}+12\Re\bigl(\Lam_{1111}-\Lam_{111\bar{1}}\bigr)\,,
\\
\lam_{2223}&=&3\Im\bigl[2\Lam_{11\bar{1}\bar{2}}-4\Lam_{1112}-\Lam_{111\bar{2}}+3\Lam_{112\bar{1}}\bigr]\,,
\\
\lam_{2224}&=&3\Re\bigl[2\Lam_{11\bar{1}\bar{2}}+4\Lam_{1112}-\Lam_{111\bar{2}}-3\Lam_{112\bar{1}}\bigr]\,,
\\
\lam_{2233}&=&4\Lam_{12\bar{1}\bar{2}}-2\Re\bigl[\Lam_{11\bar{2}\bar{2}}+6\Lam_{1122}+3\Lam_{112\bar{2}}-3\Lam_{122\bar{1}}\bigr]\,,
\\
\lam_{2234}&=&2\Im\bigl(\Lam_{11\bar{2}\bar{2}}-6\Lam_{1122}+3\Lam_{122\bar{1}}\bigr)\,,
\\
\lam_{2244}&=&4\Lam_{12\bar{1}\bar{2}}+2\Re\bigl[\Lam_{11\bar{2}\bar{2}}+6\Lam_{1122}-3\Lam_{112\bar{2}}-3\Lam_{122\bar{1}}\bigr]\,,
\\
\lam_{2333}&=&3\Im\bigl[2\Lam_{12\bar{2}\bar{2}}+4\Lam_{1222}+3\Lam_{122\bar{2}}-\Lam_{222\bar{1}}\bigr]\,,
\\
\lam_{2334}&=&\Re\bigl[2\Lam_{12\bar{2}\bar{2}}-12\Lam_{1222}-3\Lam_{122\bar{2}}+3\Lam_{222\bar{1}}\bigr]\,,
\\
\lam_{2344}&=&\Im\bigl[2\Lam_{12\bar{2}\bar{2}}-12\Lam_{1222}+3\Lam_{122\bar{2}}+3\Lam_{222\bar{1}}\bigr]\,,
\\
\lam_{2444}&=&3\Re\bigl[2\Lam_{12\bar{2}\bar{2}}+4\Lam_{1222}-3\Lam_{122\bar{2}}-\Lam_{222\bar{1}}\bigr]\,,
\\
\lam_{3333}&=&6\Lam_{22\bar{2}\bar{2}}+12\Re\bigl(\Lam_{2222}+\Lam_{222\bar{2}}\bigr)\,,
\\
\lam_{3334}&=&6\Im\bigl(2\Lam_{2222}+\Lam_{222\bar{2}}\bigr)\,,
\\
\lam_{3344}&=&2\Lam_{22\bar{2}\bar{2}}-12\Re\Lam_{2222}\,,
\\
\lam_{3444}&=&-6\Im\bigl(2\Lam_{2222}-\Lam_{222\bar{2}}\bigr)\,,
\\
\lam_{4444}&=&6\Lam_{22\bar{2}\bar{2}}+12\Re\bigl(\Lam_{2222}-\Lam_{222\bar{2}}\bigr)\,.
\eeqa
\endgroup

\end{appendices}

\end{document}